\documentclass{jfm}
\usepackage{graphicx}
\usepackage{epstopdf, epsfig}

\usepackage{xcolor}
\usepackage{amsmath}
\usepackage{pdflscape}
\usepackage{array}
\usepackage{float}
\usepackage{comment}
\usepackage{multirow}
\excludecomment{toexclude}

\shorttitle{Revisiting wind wave growth with fully-coupled direct numerical simulations}
\shortauthor{Jiarong Wu, Stéphane Popinet, and Luc Deike}

\title{Revisiting wind wave growth with fully-coupled direct numerical simulations}

\author{Jiarong Wu\aff{1}, Stéphane Popinet\aff{2}, and Luc Deike\aff{1,3,\corresp{\email{ldeike@princeton.edu}}}}

\affiliation{\aff{1}Department of Mechanical and Aerospace Engineering, Princeton University, Princeton, NJ 08544, USA
\aff{2}Institut Jean Le Rond d’Alembert, CNRS UMR 7190, Sorbonne Université, Paris 75005, France
\aff{3}High Meadows Environmental Institute, Princeton University, Princeton, NJ 08544, USA}

\begin{document}

\maketitle

\begin{abstract}
We investigate wind wave growth by direct numerical simulations solving for the two phase Navier-Stokes equations. We consider ratio of the wave speed $c$ to wind friction velocity $u_*$ from $c/u_*=$ 2 to 8, i.e. in the slow to intermediate wave regime; and initial wave steepness $ak$ from 0.1 to 0.3; the two being varied independently. The turbulent wind and the travelling, nearly monochromatic waves are fully coupled without any subgrid scale models. The wall friction Reynolds number is 720. The novel fully-coupled approach captures the simultaneous evolution of the wave amplitude and shape, together with the underwater boundary layer (drift current), up to wave breaking. The wave energy growth computed from the time-dependent rms surface elevation is in quantitative agreement with that computed from the extracted surface pressure distribution, which confirms the leading role of the pressure forcing for finite amplitude gravity waves. The phase shift and the amplitude of the principal mode of surface pressure distribution are systematically reported, to provide direct evidence for possible wind wave growth theories. Intermittent and localised airflow separation is observed for steep waves with small wave age, but its effect on setting the phase-averaged pressure distribution is not drastically different from that of non-separated sheltering. For the momentum and energy fluxes, we find that the wave form drag force is not a strong function of wave age but closely related to wave steepness. The time evolution of  the rms steepness and the wave form drag suggests that there is an effect of the history of wind wave coupling, with waves of different initial steepnesses $ak$ resulting in different wave form drag values at the same instantaneous $a_{rms}k$ later, which is due to the different wave crest shape and other complex coupling effects. The normalised wave growth rate we obtain agrees with previous experimental and numerical studies. We make an effort to clarify various commonly-adopted underlying assumptions, and to reconcile the scattering of the data between different previous theoretical, numerical, and experimental results, as we revisit this longstanding problem with new numerical evidence. 
\end{abstract}


\section{Introduction}
\subsection{Motivation}
Wind waves, i.e. waves forced by local wind, play an active role in many air-sea interaction processes \citep[][]{SULLIVAN2010,CAVALERI2012,DEIKE2022}. The growth of waves under wind forcing, however, is still an area with open questions, in terms of the exact mechanism responsible for wave growth. A number of theories \citep[][]{JEFFREYS1925,MILES1957,BELCHER1993} of varying complexity have been proposed over the years (see \citet[][]{JANSSEN2004} for a review) but their applicability is unclear due to lack of direct empirical evidence. Field campaigns \citep[][]{SNYDER1981,DONELAN2006} and laboratory scale experiments \citep[][]{PEIRSON2008,GRARE2013a,SHEMER2019,BUCKLEY2020a} have reported growth rates that can scatter by an order of magnitude, and sometimes largely deviate from the theoretical predictions \citep[see][]{PEIRSON2008}. Since the wind forcing forms the basic source term for any operational wave model \citep{JANSSEN2004}, it is important to continue to improve our physical understanding of the dynamic processes controlling the wave growth rate in different wind-wave regimes.

\subsection{Problem formulation}
\begin{figure}
    \centering
    \vspace{0.5cm}
    \includegraphics{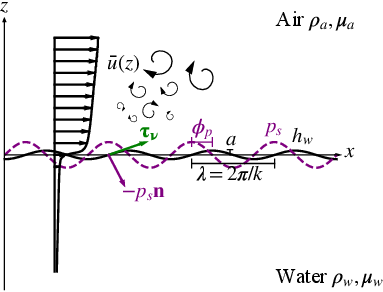}
    \caption{A sketch of the wind-wave problem. The surface stress consists of the normal pressure stress ($p_s \boldsymbol{n}$), and the viscous stress $\boldsymbol{\tau}_{\nu}$. The correlation of the surface pressure $p_s$ (purple dotted line) with the surface elevation slope $\partial h_w/\partial x$ is generally thought to be the major contribution to the wave growth (see (\ref{eqn:intro1})). In this paper we consider wind blowing in $x$ direction, and therefore no misalignment effect is discussed. The wind blows from left to right, and the maximum of the pressure distribution is on the windward face for slow moving waves. The phase shift $\phi_p$  denotes the phase lag of the pressure maximum to the wave crest.}
    \label{fig:diagram}
\end{figure}

The dynamics of the wind wave interaction is a coupled two-phase flow, as sketched in figure \ref{fig:diagram}. The wind (of density $\rho_a$) blows across a moving wavy water surface $h_w (x,y,t)$ (of density $\rho_w$), and the structure of the atmospheric turbulent boundary layer is altered. The resulting wave coherent surface wind stress in turn transfers energy into the waves. 
The wind stress at the surface consists of two parts, the viscous stress ($\boldsymbol{\tau}_{\nu}$) mostly in the tangential direction, and the pressure stress ($p_s\boldsymbol{n}$) in the normal direction, see figure \ref{fig:diagram}. 
It has been generally agreed on that for gravity waves, the wave growth mostly results from the work done by the surface air pressure, although the wave coherent viscous stress can play a part at low steepness and gravity-capillary waves \citep[][]{PEIRSON2008,BUCKLEY2020a} and force the underlying current \citep[][]{WU1968,LIN2008,WU2021}. With this widely adopted assumption (which we will test explicitly in this paper), the energy input rate can be written as \citep{GRARE2013a} 
\begin{equation}\label{eqn:intro1}
    S_{in} \approx \langle -p_s\boldsymbol{n}\cdot \boldsymbol{u_s} \rangle \approx c\langle p_s\frac{\partial h_w}{\partial x}\rangle
\end{equation}
where $S_{in}$ denotes the wave-averaged rate of energy input flux. The angular brackets denote averaging over one wavelength, and $\boldsymbol{u_s}$ is the surface water velocity. The part of $\boldsymbol{u_s}$ that is correlated to the pressure is by linear approximation the vertical wave orbital velocity $w_{orbit} = -c(\partial h_w / \partial x)$, with $c$ the wave phase speed. The derivative in $x$ assumes that the waves are predominantly 2D and travelling in the $x$ direction. Note that the average also defines the wave form drag $F_p$:
\begin{equation}
F_p = \langle p_s\partial h_w/\partial x\rangle
\end{equation} 
similar to the concept of the form drag of a blunt body.


Based on (\ref{eqn:intro1}), the key to determine the rate of energy input is the correlation between the surface pressure profile and the surface slope. Experimental measurements \citep[][]{PLANT1982,PEIRSON2008,GRARE2013a,BUCKLEY2020a,FUNKE2021} have directly or indirectly estimated this correlation (more on the experimental methods in \S\ref{sec:intro_experiments}). It is also a framework that most theoretical works have adopted. 

\subsection{A brief review on the representation of surface pressure in wind wave growth theories} \label{sec:intro_theory}

We first present a brief review of some of the theories developed over the years to describe wind wave growth, and how they have affected the representation and comparison of experimental data. 

\citet[][]{JEFFREYS1925} was the earliest to propose what is now called the `sheltering hypothesis', where the surface pressure is assumed to be $90^{\circ}$ out of phase with the surface, i.e. in phase with the slope,
\begin{equation}\label{eqn:Jeffrey_p}
	p_s = s_z\rho_a (U_z-c)^2 \frac{\partial h_w}{\partial x},
\end{equation}
where $s_z$ is the non-dimensional sheltering coefficient, and $U_z$ a reference velocity at a given height $z$. The choice of the reference velocity is not specified, and (\ref{eqn:Jeffrey_p}) can be interpreted as a scaling analysis.
The energy input rate $S_{in}$ follows (\ref{eqn:intro1}) and reads
\begin{equation}\label{eqn:Jeffrey_Sin}
	S_{in} = \frac{1}{2}\rho_a s_z (ak)^2c(U_z-c)^2,
\end{equation}
assuming that the surface elevation has the sinusoidal form $h_w = a\cos(kx)$. The viscous stress input was assumed to be negligible compared to the pressure input. Jeffrey's original idea is that the airflow is separated behind the wave crest, and therefore, his theory is not limited to small amplitude waves. 


\citet[][]{MILES1957} proposed the critical layer theory through a linear stability analysis. The airflow is assumed to be inviscid and laminar, and as a result of that assumption, the forcing comes solely from the pressure. The shifted pressure profile is assumed the complex form
\begin{equation} \label{eqn:Miles_p}
	p_s = (\alpha + i\beta)\rho_aU_{ref}^2 k h_w
\end{equation}
while the surface elevation $h_w$ is
\begin{equation}
	h_w = ae^{i(kx-\omega t)}
\end{equation}
Again $U_{ref}$ is an arbitrarily chosen reference velocity. The energy input, however, was not computed from (\ref{eqn:intro1}), but from a change to the complex wave phase speed $c$ through the boundary condition at the interface,
\begin{equation}
    c = c_0 + \frac{1}{2}\frac{\rho_a}{\rho_w}(\alpha + i\beta)(U_{ref}/c_0)^2.
\end{equation}
where $c_0$ is the phase speed of a free surface gravity wave.
The wave energy rate of change $dE/dt$ (or $S_{in}$) is normalised by the wave angular frequency $\omega$ and the wave energy $E$ in order to yield the growth rate form of
\begin{equation} \label{eqn:Miles_dEdt}
\gamma = \frac{1}{\omega E}\frac{dE}{dt} = \frac{S_{in}}{\omega E} \approx 2\Im(c)/\Re(c) = \beta \frac{\rho_a}{\rho_w} \left(\frac{U_{ref}}{c}\right)^2,
\end{equation}
neglecting wave dissipation by viscosity. $\Im(c)$ and $\Re(c)$ are the imaginary and the real part of $c$ respectively.
In another word, the perturbation grows exponentially under the linear stability analysis, and finding the growth rate (per radian) $\gamma$ is equivalent to finding $\beta$, the imaginary part of the surface pressure distribution. This requires solving the Rayleigh equation, and $\beta$ was found to be related to the curvature of the mean wind velocity profile at the critical height (where the wind speed equals the wave phase speed).



The applicability of the critical layer theory has been questioned, as it ignores turbulence effects; for short and slow travelling waves, the critical layer is very close to the water surface \citep[][]{BELCHER1993, JANSSEN2004}, where viscous effect might be important; it also does not capture the effect of finite amplitude or steep waves \citep[][]{PEIRSON2008}. As an improvement to Miles' theory, \citet[][]{BELCHER1993} and \citet[][]{BELCHER1999} incorporated the turbulence's effects and proposed the non-separated sheltering mechanism. The turbulent boundary layer is divided into the inner surface layer, the stress surface layer, the middle layer and the outer layer based on the asymptotic structure of the flow. The surface pressure is
\begin{equation}\label{eqn:Belcher_p}
	p_s = (-1 + i\frac{u_*^2}{U_m^2}\beta)\rho_a U_m^2 kh_w,
\end{equation}
where $U_m$ is the middle layer velocity, and $\beta$ was attributed to a few different mechanisms. Since only the turbulent stress is considered, which goes to zero at the surface, the energy input is by construction only done by the surface pressure.

All the above theories have attributed the energy input to the surface pressure forcing. What (\ref{eqn:Jeffrey_p}), (\ref{eqn:Miles_p}), (\ref{eqn:Belcher_p}) have in common is a phase shifted pressure profile, and the amplitude of the pressure profile given by $\rho_a$ times some reference velocity $U^2_{ref}$ ((\ref{eqn:Jeffrey_p}) can be written as 
$p_s = is_z\rho_a (U_z-c)^2 kh_w$, and the sheltering coefficient $s_z$ is equivalent to $\beta$ if $U_z-c = U_{ref}$). Understanding what controls the phase shift and the reference velocity in various regimes, however, is no easy work, and depends on the specific proposed mechanism, as well as the mean wind velocity profile. 

\subsection{Connecting theoretical growth rate and observations} \label{sec:intro_experiments}

Experimental measurements of the input rate $S_{in}$ have followed different approaches.
One option is to measure the correlation $\langle p_s \partial h_w/\partial x \rangle$ in (\ref{eqn:intro1}) by simultaneous measurement of the pressure and the surface elevation \citep[][]{SNYDER1981,DONELAN2006,GRARE2013a}. Direct measurement of the surface pressure requires complex wave following pressure sensors, which tend to be limited in responding frequencies, and have to be placed at a certain height above the water surface, which introduces additional uncertainty \citep[][]{DONELAN2006,GRARE2009}. Alternatively \citet[][]{BUCKLEY2020a} performed PIV measurements of the air flow above the wave and estimated the pressure forcing as residual stress or from pressure reconstruction \citep[][]{FUNKE2021}.


The other option is to directly measure the wave energy growth from temporal or spatial evolution of the surface elevation \citep[][]{KAWAI1979,PEIRSON2008}. The wave energy rate of change is related to the energy input rate by
\begin{equation} \label{eqn:intro2}
    S_{in} = D + dE/dt,
\end{equation}
where $D$ is the wave dissipation term, usually estimated from the linear viscous dissipation rate \citep[][]{LAMB1993}
\begin{equation} \label{eqn:dissipation}
    D = 4\nu_w k^2 E
\end{equation}
where $\nu_w = \mu_w/\rho_w$ is the kinematic water viscosity. The dissipation term $D$ is small for relatively long waves above \textit{O}(1m) but not negligible in some lab scale experiments. This method measures $S_{in}$ without the assumption that the pressure forcing is the dominant contribution \citep[][]{PEIRSON2008}. The difficulty then resides in measuring the small fraction of change in the wave amplitude given the small values of the wave growth due to the small density ratio $\rho_a/\rho_w$. Uncertainties in the dissipation rate also remains, due to the role of parasitic capillary waves or micro-breaking that can dominate over the viscous dissipation especially in finite amplitude cases \citep[][]{GRARE2013a}.


The experimental and field measurements of the energy input rate $S_{in}$ have shown a reasonable agreement with (\ref{eqn:Miles_dEdt}), adopting the air friction velocity $u_*$ as the reference velocity \citep[][]{PLANT1982}. The definition of $u_*$ is based on the total downward momentum transfer and carries some uncertainty itself. There are other choices of the reference velocity, and therefore other representations of $\gamma$. For example, \citet[][]{DONELAN2006} adopted the sheltering hypothesis and found that using the wind velocity at half the wavelength $U_{\lambda/2} - c$ in (\ref{eqn:Jeffrey_Sin}) best collapsed their data. 

To summarise, the experimental uncertainties, together with the indirect nature of the estimations of the energy input rate make it difficult to directly verify a specific growth mechanism. A direct connection to the various theories would require knowledge of not just the wave-averaged quantity $S_{in}$, but also the phase resolved pressure profile $p_s$. Few experimental works (\citet[][]{BANNER1990,DONELAN2006, GRARE2009} to our knowledge) have discussed the pressure profile itself, due to the difficulty of pressure measurement.

Numerical simulations have much to offer in this regard, and can focus on either the wind or the wave side. Simulations focused on the turbulent airflow over a wavy boundary (stationary or with prescribed wave motion) have been conducted using both direct numerical simulations (DNS) \citep[e.g.][]{SULLIVAN2000,KIHARA2007a,YANG2010,DRUZHININ2012a} and large eddy simulation (LES) \citep[e.g.][]{YANG2013,SULLIVAN2014,SULLIVAN2018a,SULLIVAN2018b}. They provide detailed information about the wave induced perturbation and stresses, and the wave growth is inferred from (\ref{eqn:intro1}). DNS does not require subgrid scale models but is limited by the high computational cost associated with high Reynolds number. 
Wall modelled LES, on the other hand, is able to simulate much higher Reynolds number flows, but the subgrid scale models for wave drag is still under development \citep[][]{DESKOS2021, AIYER2022}. Most importantly, wall modelled LES, by design, does not offer enough insight into the dynamics of wave growth since the wall models assume knowledge of this process \citep[][]{PIOMELLI2002}. Wall resolved LES, which takes a middle ground, has been applied to the study of a broadband wave field growth \citep[][]{YANG2013}, but is also restricted in the Reynolds number similar to DNS. Simulations focused on the wave evolution usually simplify the wind effects into a forcing at the water top boundary, either as solely a phase-shifted pressure distribution \citep[][]{FEDOROV1998,ZDYRSKI2020}, or as both pressure and viscous shear stress distribution \citep[][]{TSAI2013}. This requires the stress distribution as prior knowledge, which as we have discussed, is far from understood.

The importance of air flow separation and breaking waves on the form drag has long been recognized \citep{BANNER1990,BANNER1998} and simulations with prescribed wave shapes based on laboratory work have attempted to quantify this effect \citep{SULLIVAN2018a,SULLIVAN2018b}. In this regime, the waves are highly nonlinear. \citet{YANG2010} have found that nonlinearity can have an appreciable impact on wave form drag and thus the growth rate, which calls for the inclusion of `realistic wave dynamics' rather than ideal wave shape, and `coupled simulation of wind and wave motions'. However, to this date, the majority of the numerical works on wind waves are limited to one side of the problem and not coupled. To our knowledge, the only numerical works where both the wind and the growth of the surface waves are directly resolved are in the context of the very initial wave generation \citep[][]{LIN2008,KOMORI2010,TEJADA-MARTINEZ2020,LI2022}.

What distinguishes this work from previous numerical works is therefore the fully-coupled approach for finite amplitude waves. We extend our earlier 2D study with linear-shearing laminar wind forcing \citep[][]{WU2021} to a 3D turbulent boundary layer wind forcing. We use a volume of fluid (VoF) method to reconstruct the interface and access the wave growth, including the case of steep waves. We can access the wave growth from directly observable wave evolution, in addition to inferring it from the pressure-slope correlation. This allows us to verify the assumption (\ref{eqn:intro1}) that $S_{in}$ mostly results from the pressure stress. We also discuss the spatial structure of the pressure field and phase shift with the wave profile.
We study independently the effects of two key parameters, the wave steepness $ak$ and the ratio between the wave phase speed and the wind friction velocity $c/u_*$ (referred to as wave age in wind wave literature). In experiments, the two parameters are connected by the fetch-limited relation, and therefore their respective effects are hard to separate. This numerical approach also allows us to expand the parameter range to steeper and even breaking waves, and study the effect of airflow separation and breaking in this regime, while the wind and the waves are fully coupled.



The paper is structured as follows. In \S\ref{sec:numerics} we introduce the numerical setup. In \S\ref{sec:evolution} we qualitatively describe the time evolution of the fully coupled wind-wave system, and the mean profiles in the air and in the water. In \S\ref{sec:qualitative} we define the wave averaged quantities of interests: the wave energy, and the momentum and energy fluxes. We cross-check the wave growth obtained from wave surface elevation and from the pressure-slope correlation. We also discuss the time evolution of the wave form drag together with geometric features of the waves. In \S\ref{sec:pressure_distribution} we present the surface pressure distribution (phase shift and amplitude) for different $c/u*$ and initial $ak$ values. In \S\ref{sec:bulk_parameter} we discuss the scaling of the wave form drag, and the energy input rate with $c/u_*$ and $ak$. We compare with previous data sets and discuss the implications for possibly applicable theories. 

\section{Direct numerical simulation of fully coupled wind and waves} \label{sec:numerics}
We present direct numerical simulations of fully coupled wind forced water waves. We solve the two-phase Navier-Stokes equations with the Basilisk solver \citep{POPINET2009,POPINET2015,POPINET2018,FUSTER2018}, with a momentum conserving scheme \citep{ZHANG2020} and a geometric VoF method to reconstruct the interface. We use adaptive mesh refinement (AMR) which allows us to reduce the computational cost when solving such a multi-scale problem. The methods have been extensively validated and applied to wave breaking \citep[][]{DEIKE2015,DEIKE2016,MOSTERT2020,MOSTERT2022a}, two-phase turbulent flow \citep[][]{RIVIERE2021,PERRARD2021,FARSOIYA2021}, and atmospheric turbulent boundary layer \citep[][]{VANHOOFT2018}.

\subsection{Governing equations}
We solve the incompressible Navier Stokes equations:
\begin{align}
\frac{\partial\mathbf{\rho}}{\partial t} + \mathbf{\nabla} \cdot(\rho \mathbf{u})=&\;0 \label{eqn:NS1}\\ 
\rho(\frac{\partial\mathbf{u}}{\partial t} + (\mathbf{u}\cdot\mathbf{\nabla})\mathbf{u}) =& -\mathbf{\nabla} p +\mathbf{\nabla}\cdot(2\mu\mathbf{D}) + \sigma\kappa\delta_{S}(\mathbf{x} - \mathbf{x_{\mathcal{F}}}) \mathbf{n} \label{eqn:NS2}\\
\nabla \cdot \mathbf{u} =& \; 0 \label{eqn:NS3}
\end{align} 
An additional scalar field representing the volume fraction of one of the two phases is introduced as $\mathcal{F}(x,y,z,t)$. The physical properties (i.e. density and the viscosity) are $\mathcal{F}$ weighted averaged of the densities and the viscosities of the water and air phases:
\begin{equation}
    \rho = \mathcal{F}\rho_w+(1-\mathcal{F})\rho_a, \;\; \mu = \mathcal{F}\mu_w+(1-\mathcal{F})\mu_a
\end{equation}
This together with (\ref{eqn:NS1}) - (\ref{eqn:NS3}) constitute the governing two-phase Navier-Stokes equations we numerically solve for. The $\mathcal{F}$ field evolves based on the continuity equation. A momentum-conserving scheme is implemented, and mass is well conserved with an error typically below 0.01\% \citep{MOSTERT2022a}.

\subsection{Numerical setup}
\begin{figure}
    \centering
    \vspace{0.5cm}
    \includegraphics{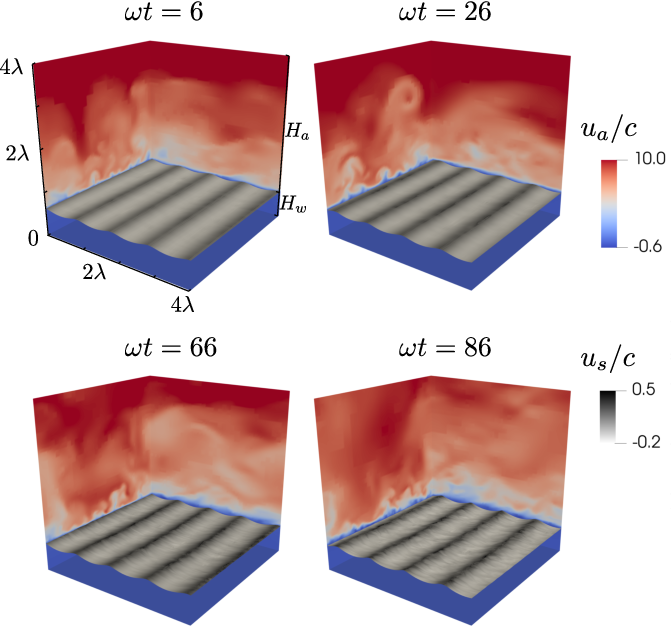}
    \caption{Snapshots of the air side turbulent boundary layer and the evolving waves, for the strongest forcing case ($c/u_*=2, ak=0.2$). There are four waves in the computational domain, and the height of the water and the half channel height for the air are shown. The colours indicate the instantaneous horizontal wind velocity $u_a$, and the surface water velocity $u_s$, respectively. The waves grow in amplitude and become short crested, which is a characteristic of wind waves. At later stage, the waves also appear to be three-dimensional because of the development of an underwater turbulent boundary layer.}
    \label{fig:snapshots}
\end{figure}

The computation domain is of size $L_0\times L_0 \times L_0$, with four waves in the $x$ direction of wavelength $\lambda = L_0/4$ (wave number $k=2\pi/\lambda=8\pi/L_0$). The depth of the resting water is $H_w = L_0/{2\pi}$, while the height of the airflow is $H_a = L_0(1-1/{2\pi})$ (see figure \ref{fig:snapshots}). The top and the bottom are both free slip boundary conditions, while the front and back, left and right are periodic boundary conditions.

We initialise the wave shape with a given surface elevation function $h_w(x,y,z, t=0)$, in this case chosen to be a third order Stokes wave shape similar to that used in \citet{WU2021}. The initial steepness $ak$ ranges from 0.1 to 0.3. The $\mathcal{F}(x,y,z, t=0)$ field is then initialised on a discretised grid based on the sign of $y-h_w(x,y,z, t=0)$. $\mathcal{F} = 1$ for the water phase ($y-h_w(x,y,z)<0$) and $\mathcal{F} = 0$ for the air phase ($y-h_w(x,y,z)>0$).

During the turbulence precursor preparation stage, the waves are kept stationary by setting $\mathcal{F}\mathbf{u}=\mathbf{0}$ at each time step. This configuration is equivalent to a turbulent boundary layer over stationary bumps.
We force the turbulence with a pressure gradient (similar to a canonical channel flow), which sets the \textit{nominal} friction velocity $u_*$ (i.e. total wall stress $\tau_{0}$)
\begin{equation} 
    \tau_{0} = \rho_{a} u_*^2 = H_a\frac{\partial{p}}{\partial{x}}.
\end{equation}
The friction Reynolds number is defined as $Re_{*} = \rho_{a}u_*H_a/\mu_{a}$ and set to $720$ for all cases. Notice that the height of the airflow is set to more than three times the wavelength $\lambda$, so that the effect of the top boundary is minimised. The physically more relevant Reynolds number is the one based on the wavelength\begin{equation}\label{eqn:friction_Re}
    Re_{\lambda} = \frac{\rho_{a}u_*\lambda}{\mu_{a}},
\end{equation}
which is 214 (the ratio of the wave and the boundary layer length scales, equivalently $k\delta_{\nu}=0.029$). We use adaptive mesh with a maximum refinement level 10 (see appendix \ref{sec:appendix1} for a detailed description of the adaptive mesh refinement feature), which means that the smallest cell size is $\Delta=L_0/2^{10}$. There are around $1.8\times10^7$ grid cells in a typically simulation case, which is less than 2\% of the uniform spaced grid of the same resolution ($1024^3\approx1.07\times 10^9$).
We have validated the solver against a canonical flat wall case with $Re_*=180$ \citep[][]{KIM1987} (see appendix \ref{sec:appendix1} for details). The mean wind velocity profile of such a channel flow follows the law of the wall, and is similar to that of laboratory wind wave experiments \citep[e.g.][]{BUCKLEY2020a}.

After the turbulence precursor reaches a statistically steady state, the waves are released at $t=0$ (meaning that there is no manually setting $\mathcal{F}\mathbf{u}=\mathbf{0}$ anymore, and the initial orbital velocity is added), and travel with a phase speed given by the free surface dispersion relation 
\begin{equation}
c = \sqrt{g/k+\sigma k/\rho_{w}},
\end{equation}  
where $g$ is the gravitational acceleration and $\sigma$ is the surface tension. The orbital velocity is initialised with the corresponding velocity field of the third order Stokes wave \citep[see][]{WU2021}.

Since we initialise the waves with a solution of the free surface gravity wave equation, we expect the flow field to self-adjust under wind forcing during the very early stage of the simulation. The turbulent boundary layer also goes through a relaxation period when the near-wall flow adjusts to the moving boundary. We define an eddy turnover time scale $T_e = 2\lambda / \langle u \rangle (z=\lambda)$, where $\lambda$ is the wavelength and $\langle u \rangle (z=\lambda)$ is the mean horizontal velocity at vertical height $z=\lambda$. Physically it is the time scale for an eddy of size comparable to the wavelength $\lambda$ to reach equilibrium with the changing flow boundary condition.  Based on both the evolution of the wind stress and the mean profiles, we observe that the relaxation period last for about $4T_e$, therefore, the data of the first $4T_e$ are not included in the physical analysis. We note that any choices of initialisation will present certain limitations as there is no exact solution of the full two-phase turbulent problem we can use to start the simulation. After the initialisation, the waves and the turbulence interact in a fully coupled way without any prescribed interfacial conditions. The whole simulation is transient by nature, meaning that the wave amplitude changes with time, despite over a much longer time scale than both the turbulence time scale and the wave period.

The non-dimensional numbers relevant for the waves are
\begin{equation}
    Bo = \frac{(\rho_{w} - \rho_{a})g}{\sigma k^2}, \;\;
    Re_w = \frac{\rho_{w}c\lambda}{\mu_{w}} \label{eqn:wave_Re}.
\end{equation}
In all the cases presented in this paper, the Bond number $Bo=200$ so that the waves are in the gravity wave regime, and we have verified that further increasing $Bo$ does not affect the results presented here (see appendix \ref{sec:appendix1}). The density ratio $\rho_a/\rho_w$ is set to air-water conditions $1/850$, while the viscosity ratio $\mu_a/\mu_w$ is always larger than 50 and is adjusted to set the air friction Reynolds number $Re_{\lambda}$ (\ref{eqn:friction_Re}) and the wave Reynolds number $Re_{w}$ (\ref{eqn:wave_Re}) independently. 
The wave Reynolds number is kept at $Re_w \approx 10^5$. Note that the value of $Re_w$ gives the linear dissipation rate (per radian) due to viscosity $\gamma_d$ \citep{LAMB1993}
\begin{equation} \label{eqn:gamma_d} 
    \gamma_d = -4\nu_wk^2/\omega = \frac{8\pi ck}{Re_w}/\omega =  8\pi/Re_w
\end{equation}
and $D = \gamma_d \omega E$ (equivalent to (\ref{eqn:dissipation})).

Notice that the velocity ratio (wave age) $c/u_*$ is varied by changing $c$, while keeping $u_*$ constant, independently of the steepness $ak$. This configuration allows to resolve the turbulent air flow and capture the wave growth for $c/u_*$ ranging from 2 to 8 and $ak$ from 0.1 to 0.3. Table 1 summarises the simulation conditions, together with the characteristic length scales of the turbulence $\delta_\nu$ and the capillary length $l_c$, relative to wave number $k$ and to the smallest grid size $\Delta$.

\begin{table}
  \begin{center}
\def~{\hphantom{0}}
  \begin{tabular}{ccccccc}
  $ak$        & $c/u_*$ & $k\delta_{\nu}$* & $kl_c$ & $a/\Delta$ & $\delta_{\nu}/\Delta$ & $k\Delta$  \\[3pt]
   0.10 & 2,4,6,8 & \multirow{4}{*}{0.029} & \multirow{4}{*}{0.44} & 4.1 &  \multirow{4}{*}{1.2} & \multirow{4}{*}{0.025}         \\
   0.15 & 2,3,4,6,8 & & & 6.1 & & \\
   0.20 & 2,4,6,8 & & & 8.1 & &\\
   0.25 & 2,4,6,8 & & & 10.2 & &\\
   0.30 & 2   &&& 12.2 & &\\
  \end{tabular}
   
  \caption{A table of controlling parameters $ak$ and $c/u_*$, and relevant length scales. The third and fourth columns are the viscous wall unit $\delta_{\nu} = \nu_a/u_*$ and the capillary length scale $l_c = 2\pi\sqrt{\sigma/(\rho_w - \rho_a)g}$ relative to $1/k$ respectively, showing the physical relevance of the parameters. They are controlled by $\text{Re}_{*}=720$ and $\text{Bo}=200$ and kept constant. The last three columns are $a$, $\delta_{\nu}$ and $k$ relative to the smallest grid size $\Delta$, showing the numerical resolution. In the simulations, $\Delta = L_0/2^{N}$, where $N=10$ is the maximum refinement level of the octree adaptive grid. *For wall modelled LES, the roughness length $kz_0$ is usually reported instead of $k\delta_{\nu}$. If we use the $z_0 = 0.11\nu_a/u_* = 0.11\delta_{\nu}$ conversion for flat smooth surface, $kz_0 = 0.003$. Also notice that these length scales are not changed when we change $c/u_*$ because $k$ is fixed, in contrast to the realistic situation, where wavenumber $k$ is smaller for fast moving waves.}
  \label{tab:length_scale}
  \end{center}
\end{table}


\section{Evolution of the fully coupled wind-wave system}\label{sec:evolution}

\begin{figure}
    \centering
    \vspace{0.5cm}
    \includegraphics{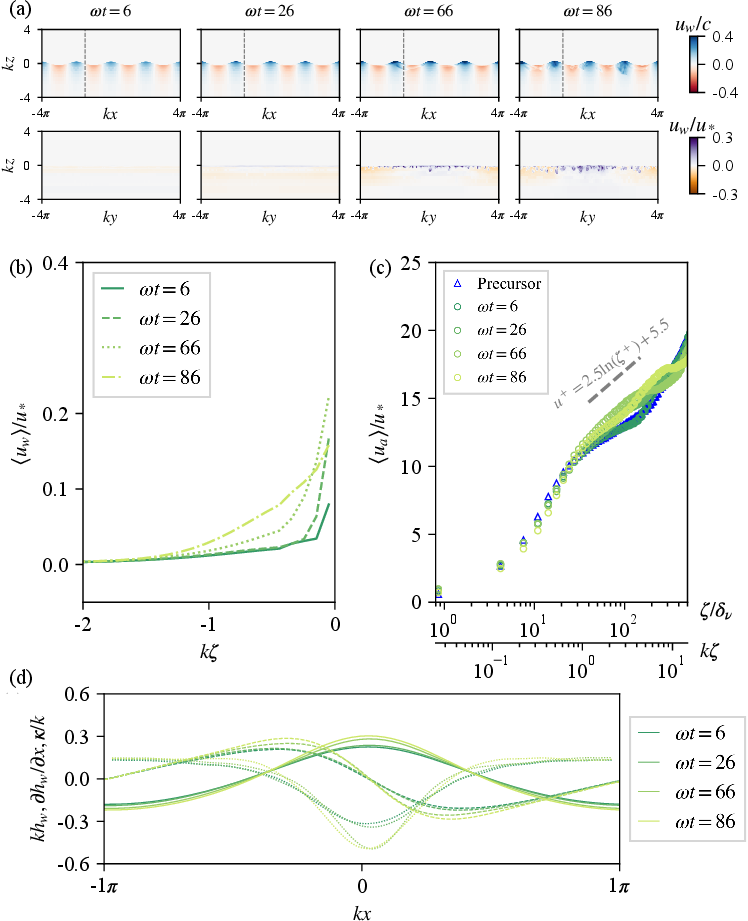}
    \caption{Simultaneous development of the waves and the associated air and water side boundary layers, for the strongest wind forcing case ($c/u_*=2, ak=0.2$), shown at four representative time. (a) The first row shows the instantaneous horizontal velocity normalised by wave phase speed $c$ in the $x-z$ plane. The horizontal velocity is the wave orbital velocity plus the drift layer. The second row also shows the instantaneous horizontal velocity $u$, but normalised by the wind friction velocity $u_*$ and in the $y-z$ plane instead (taken at the $x$ location indicated by the grey dotted line in the first row). (b) The time evolution of the average vertical profile for the underwater boundary layer. The wave-following $\zeta$ coordinate is defined in appendix \ref{sec:appendix0_2}. (c) Time evolution of the mean wind velocity profiles, for the turbulence precursor and at later times with moving waves, in the same wave-following coordinate. The $x$-axis shows the vertical $\zeta$ coordinate normalised by the viscous wall unit $\delta_{\nu}$ and the wavenumber $k$ respectively. The ratio of $k\delta_{\nu}=0.029$ is fixed in all the cases.  (d) The wave shape time evolution. The solid lines show the spanwise ($y$ direction) averaged wave shape $h_w(x,t)$; the dashed lines show the horizontal gradient $\partial h_w(x,t)/ \partial x$; the dotted lines show the curvature $\kappa$ divided by wavenumber $k$.}
    \label{fig:coupling_CU2}
\end{figure}

\begin{figure}
    \centering
    \vspace{0.5cm}
    \includegraphics{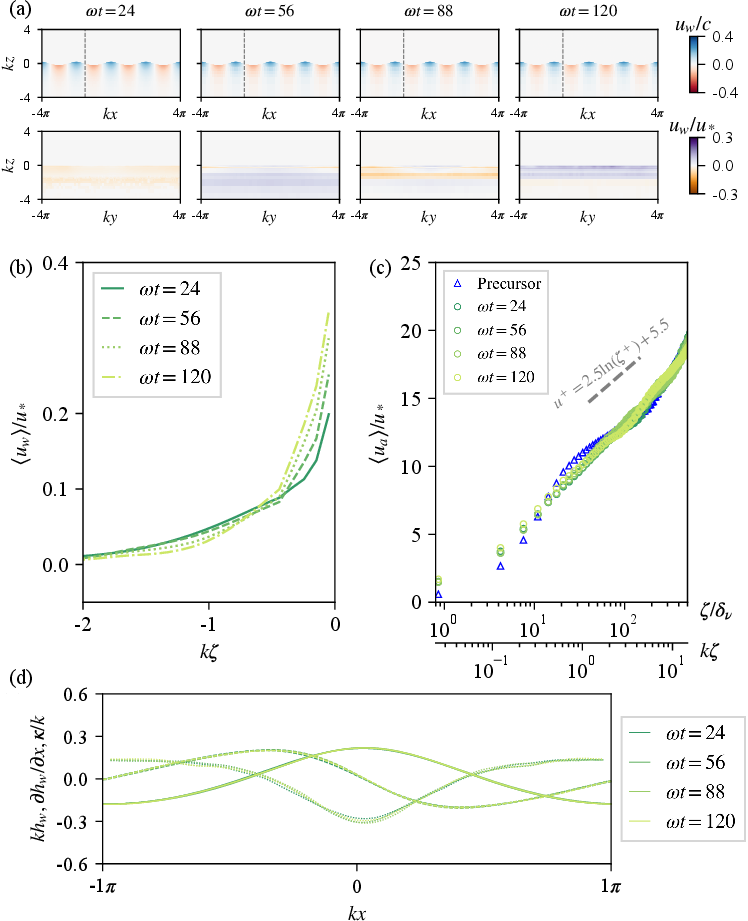}
    \caption {The same as figure \ref{fig:coupling_CU2} but for the $c/u_*=8$, $ak=0.2$ case. The underwater turbulent boundary layer development is suppressed, and the surface drift is able to reach a higher value because of this suppression. The air side turbulent boundary layer mean profile is more steady.}
    \label{fig:coupling_CU8}
\end{figure}

Figure 2 shows qualitatively the air side turbulent boundary layer. It also shows the wave surface evolving due to the turbulence forcing, with growth and steepening as the wind keeps blowing. The waves are narrow banded for most cases, as the development of higher frequency ripples and 3D structure only occurs at later times, while the downshift of peak frequency is constrained by the periodic boundary condition. However, the wave shape changes and becomes short-crested, which is a feature of wind waves. 

Since we take a fully coupled approach, there is a shear-induced drift layer development underneath the water surface while the waves develop. The waves directly feedback to the air side turbulent boundary layer as well. To illustrate the whole picture of the fully coupled system, we show all the above mentioned elements for two representative cases: one with the smallest wave age, i.e. stongest wind forcing ($c/u_*=2, ak=0.2$) in figure \ref{fig:coupling_CU2}, the other with an intermediate wave age case ($c/u_*=8, ak=0.2$) in figure \ref{fig:coupling_CU8}.

For the strongly forced case, as we can see from both the $x-z$ and $y-z$ slices figure \ref{fig:coupling_CU2}(a), the drift layer amplifies and undergoes transition to turbulence. There are small scale entraining vortices, which also cause the surface to develop 3D features (see figure \ref{fig:snapshots}, frame 3 and 4). The stream-wise vorticity here is shear-driven and not Langmuir cells, based on the small turbulent Langmuir number \citep{TSAI2013} $\text{La}=u_*/\omega k a^2>1$ for all the cases. Figure \ref{fig:coupling_CU2}(b) shows the averaged underwater velocity profiles, which start as laminar boundary layer and develop into typical turbulent boundary layer profiles at later time. For the strongest forced case, it takes about 15 wave periods for the transition to happen. Meanwhile, as shown in figure \ref{fig:coupling_CU2}(c), the air side turbulent boundary layer mean profile keeps evolving, and at some time deviates from a logarithmic profile. This could be due to the constant momentum and energy flux from the wind into the waves, meaning that the boundary layer is not in equilibrium with the evolving boundary. We also comment that we are at a relatively low Reynolds number, which might affect the logarithmic profile and its range. In figure \ref{fig:coupling_CU2}(c), we show more clearly the wave shape and amplitude change by plotting the spanwise averaged wave surface, the horizontal gradient and the normalised curvature. The asymmetric gradient curves show that the waves are becoming more short-crested over time. The curvature is defined as $\kappa=\partial^2 h_w /\partial x ^2/(1+(\partial h_w /\partial x)^2)^{3/2}$, and its value around the wave crest is another direct measure of the short-crestness for the nearly monochromatic wave train. The `natural' evolution of the waves is the key component that differentiates our numerical simulation from the previous simulations of the turbulent boundary layer over waves where the wave shape and the wave motion are prescribed.

In contrast to the strongly forced case of $c/u_*=2$, for the $c/u_*=8$ case, the growth of the wave amplitude is much slower, almost indistinguishable, and the wave shape does not change significantly. Figure \ref{fig:coupling_CU8}(b) suggests that  the transition to turbulence of the underwater drift layer is suppressed by the larger regular wave orbital velocity, which actually allows a higher drift velocity at the surface, as shown in \ref{fig:coupling_CU8}(b). It might also be related to a longer transition to turbulence time. Figure \ref{fig:coupling_CU8}(c) shows that there is less temporal variation in the spatially averaged wind velocity profile. 
The effect of $ak$ and $c/u_*$ on the mean wind profiles are further discussed in the appendix \ref{sec:appendix0_2}.

In summary, the flow is transient in the strongly forced cases, with waves growing (and becomes more short-crested as they grow), involving the turbulent boundary layers to constantly adjust with time. On the other end, at lower forcing (higher c/u*), the very slow wave growth is negligible for the air side turbulent boundary layer, while the water side boundary layer develops slowly.
The transient behaviour is also reflected in the time evolution of the wind stress, as we will discuss in \S\ref{sec:Fp_evolution}. For the underwater drift current, its development and interaction with the waves is a problem in itself. However, the drift's effect on the wave growth is secondary if not negligible.
Here, we focus on the wave growth, and content ourselves with this brief and qualitative discussion of the underwater drift layer. In the following sections, we discuss the wind stress and its relation to the wave growth.

\section{Direct observation of the wind wave growth and the surface stress}\label{sec:qualitative}

\subsection{Directly observed wave growth}\label{sec:qualitative1}

\begin{figure}
    \centering
    \includegraphics{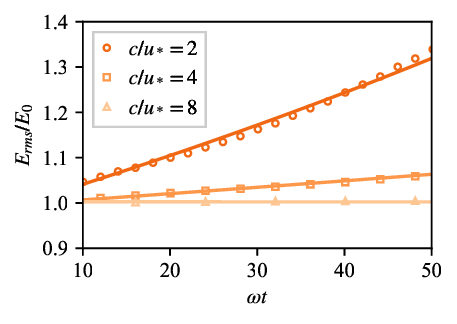}
    \caption{Wave energy normalised by initial energy $E_0 \equiv E_{rms}(t=0)$, as a function of time, directly computed from water surface height output $h_w(x,y,t)$, for three different wave $c/u_*$ = 2,4,8, and initial steepness $ak = 0.2$. The solid curves are exponential fits to the points, although we caution that the growth rates are so small that for the exponential growth cannot be distinguished definitively from a linear growth. The $c/u_*=2$ case grows the fastest while the $c/u_*=8$ case is very slowly decaying. Note that both $E_0$ and $\omega$ change with $c/u_*$ because $g$ is changed in the numerical setup (see \S\ref{sec:numerics}).}
    \label{fig:growth_compiled}
\end{figure}

We quantify the growth of the waves through the time evolution of the water surface elevation $h_w (x,y,t)$, which we use to directly compute the wave energy (neglecting the surface tension energy):
\begin{equation} \label{eqn:E_rms}
    E_{rms}(t) = \rho_w g  \langle h_w^2(x,y,t) \rangle.
\end{equation}
with the spatial wave averaging of a quantity $q$, in the $x-y$ plane, being defined as
\begin{equation} \label{eqn:<q>}
    \langle q \rangle = \frac{1}{L_0^2} \int_{-L_0/2}^{L_0/2}\int_{-L_0/2}^{L_0/2} q \; dxdy 
\end{equation}

Figure \ref{fig:growth_compiled} shows the time evolution of $E_{rms}(t)$ for three different $c/u_*$ cases, with initial wave steepness $ak = 0.2$. The smallest wave age case has the strongest wind forcing, and therefore the largest growth rate. The $c/u_* = 8$ case presents an almost exact balance between the wind input and viscous dissipation, resulting in a nearly constant wave energy with time. From this directly observed wave growth, we can measure a temporal rate of change of energy $dE/dt$ (here after we omit the subscript $rms$ for brevity).
The wave growth is rather slow, and happens over \textit{O}(10) wave periods and \textit{O}(100) to \textit{O}(1000) turbulent wall time scale $t_{\nu}=\delta_{\nu}/u_*$. This slow change in the wave energy is related to the small density ratio $\rho_a/\rho_w$, which implies weak air-water coupling; see (\ref{eqn:Miles_dEdt}).

\subsection{Wind surface stress}\label{sec:qualitative2}
\begin{figure}
    \centering
    \includegraphics{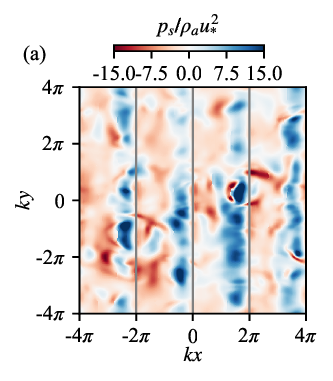}
    \includegraphics{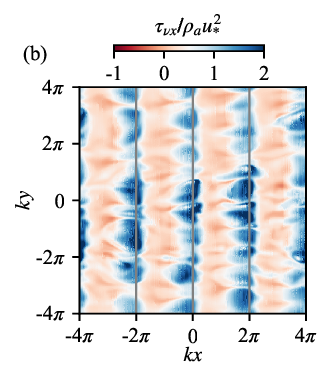}
    \includegraphics{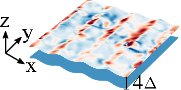}
    \caption{Instantaneous pressure (a) and the horizontal component of viscous stress (b) projected onto the wave following surface $4\Delta=0.1/k$ above the water surface, at $\omega t = 38$, for the case of $c/u_*=2$ and $a_0k=0.2$. Notice that there is one order of magnitude difference in the colour scale range. The grey lines show where the wave crests are. There are clearly wave coherent patterns.}
    \label{fig:pressure_shear} 
\end{figure}

Apart from the direct surface elevation $h_w$, we extract the surface stress from the simulation.
The wind stress \textit{at the surface} consists of two parts, the pressure variation $\boldsymbol{\tau_p}$ (i.e. drag force) and the viscous stress $\boldsymbol{\tau_\nu}$ \citep[see][]{GRARE2013a,PEIRSON2008}:
\begin{equation} \label{eqn:stresses_def}
    \boldsymbol{\tau_p} = -p_s\boldsymbol{n}, \;\;\;\;
    \boldsymbol{\tau_\nu} = \mu_a(\boldsymbol{\nabla}\boldsymbol{u_a}+\boldsymbol{\nabla}\boldsymbol{u_a}^T)\cdot \boldsymbol{n} = (\tau_{\nu x}, \tau_{\nu y}, \tau_{\nu z})
\end{equation}
where $\boldsymbol{u_a}$ is the air velocity vector, $p_s$ is the surface pressure, $\boldsymbol{n}$ is the normal vector of the water surface. 
The stresses are computed in the post-processing steps, which are independent of the computational steps. We first interpolate the velocity and pressure fields from an unstructured octree grid onto a Cartesian grid using a nearest interpolation method. The stress computation is based on the interpolated Cartesian grid. More specifically, the pressure is further interpolated onto a surface defined by $\eta + 4\Delta$, and the mean pressure along the $x$ direction is subtracted. For the shear stress, the velocity gradients are interpolated the gradients onto the same plane, while the normal vector $\boldsymbol{n}$ of the surface is constructed by the VoF method.

Figure \ref{fig:pressure_shear} shows the instantaneous stress fields projected onto the wave following surface $\eta + 4\Delta$. Since the plane is in the viscous sublayer, it is considered close enough to the actual surface that the turbulent stress can be ignored. Both the pressure and the shear stress present clear wave coherent patterns, while also having 3D structures due to the turbulence. For example, the streaks shown in figure \ref{fig:pressure_shear}(b) are about $100\delta_{\nu}$ apart, which is consistent with the typical structure of wall bounded turbulent flows. There is an-order-of-magnitude difference between the pressure and shear stress (but not their horizontal projection in (\ref{eqn:momentum_partition})). The maximum of the pressure appears on the windward face, which is left to the grey line indicating the wave crest in figure \ref{fig:pressure_shear}; this gives rise to the non-zero correlation in (\ref{eqn:intro1}). The viscous shear stress also reaches maximum near the wave crest due to the straining of the shear layer.

From the stress field we can compute the wave averaged integral quantities: the momentum flux (total stress $\tau_{total}$) and the energy flux (input rate $S_{total}$).
The total horizontal wind stress
\begin{equation} \label{eqn:momentum_partition}
    \tau_{total} = \langle \boldsymbol{\tau_p} \cdot \boldsymbol{e_x} \rangle + \langle \boldsymbol {\tau_{\nu}} \cdot \boldsymbol{e_x} \rangle = \langle p_s\frac{\partial h_w}{\partial x} \rangle + \langle \tau_{\nu x} \rangle \equiv F_p + F_s,
\end{equation}
is the sum of the form drag force per unit area $F_p$ and the averaged viscous stress in the horizontal direction $F_s$. Notice that the linear approximation ($d\eta/dx \ll 1$) is considered. 

This stress (momentum) partition is closely related to, but different from the energy input partition. The total energy input rate by the wind stress (into \textit{both} waves and underwater drift layer) is a product of the stress and the surface water velocity
\begin{equation} \label{eqn:energy_partition}
    S_{total} = \langle \boldsymbol{\tau_{total}}\cdot \boldsymbol{u_s} \rangle = \langle -p\boldsymbol{n}\cdot \boldsymbol{u_s} \rangle + \langle \boldsymbol{\tau_{\nu}}\cdot \boldsymbol{u_s} \rangle \equiv S_p + S_s.
\end{equation}
The part of $\boldsymbol{u_s}$ that correlates with the pressure is the vertical orbital velocity $w_{orbit}$, which gives (\ref{eqn:intro1}); the part of $\boldsymbol{u_s}$ that correlates with the viscous stress, however, contains both the wave horizontal orbital velocity $u_{orbit}$ and the drift velocity $u_d$:
\begin{equation}
    S_s = \langle \boldsymbol{\tau_{\nu}}\cdot \boldsymbol{u_s} \rangle \approx \langle \tau_{\nu x}{u_{sx}} \rangle = \langle \tau_{\nu x}{u_{orbit}} \rangle + \langle \tau_{\nu x}{u_{d}} \rangle \equiv S_{s,w} + S_{s,d},
\end{equation}
where $S_{s,w}$ and $S_{s,d}$ denote the energy input by the viscous shear stress into the waves and the drift respectively. The development of the drift is discussed in \citet[][]{WU2021}, and here we focus on the energy input into the waves
\begin{equation}
    S_{in} = S_p + S_{s,w} = c \langle p_s\frac{\partial h_w}{\partial x} \rangle + \langle \tau_{\nu x}{u_{orbit}} \rangle.
\end{equation}
Notice that the assumption $S_p = cF_p$ means that the pressure induced form drag contributes solely to the wave growth, while only a small variation of the viscous shear stress is correlated with $u_{orbit}$ and can contribute to the wave growth \citep[][]{PEIRSON2008}.
In other words, it is not the mean stresses but the correlated part of the stresses with the \textit{wave} surface velocity that contributes to the wave energy growth. In reality, $F_p$ and $F_s$ are of the same order of magnitude, but $S_p$ is generally thought to play a dominant role over $S_{s,w}$ (i.e. $S_{in} \approx S_p$), as mentioned in the introduction. We will examine both the momentum and the energy partitions using the simulation data.

In this paper we refer to the form drag $F_p$ as the wave form drag, and drag coefficient as the ratio $F_p/\tau_{total}$. Note that the wave drag force in the literature sometimes refers to the effective stress that contributes to the wave growth (from the energy flux $S_{in}$, instead of the momentum flux partition), and includes the pressure and the wave coherent viscous stress \citep[][etc.]{PEIRSON2008,GRARE2013a,MELVILLE2015,BUCKLEY2020a},
\begin{equation} \label{eqn:tau_w}
    \tau_{w} = S_{in}/c = F_p + S_{w,s}/c.
\end{equation} 


\subsection{Wave energy growth rate vs pressure input rate} \label{sec:pressure_energy_comparison}
\begin{figure}
    \centering
    \includegraphics{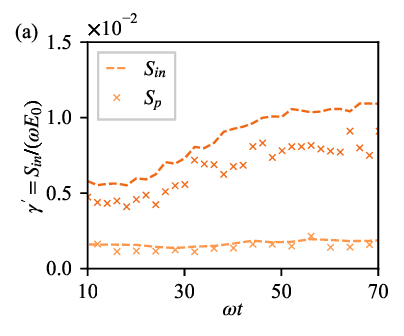} \hspace{-0.5cm}
    \includegraphics{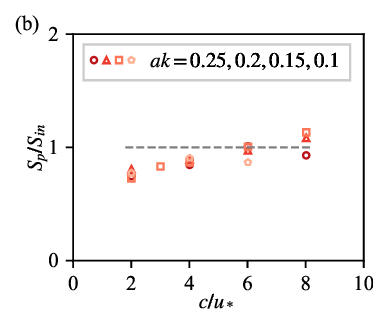}
    \caption{(a) The instantaneous pressure energy input rate $S_p=cF_p=c\langle p_s \partial h_w / \partial x\rangle$ closely follows the instantaneous wave energy growth rate (corrected with dissipation) $S_{in} = dE/dt+D$ for $c/u_*=2$ (dark orange) and $c/u_*=4$ (light orange); both are of $ak=0.2$. The curves are smoothed out using a moving window averaging. The variation in $S_p$ is mostly due to turbulence fluctuation. (b) The ratio between the averaged pressure energy input rate $S_p$ and the total input rate $S_{in} = (E(t_1)-E(t_2))/(t_1-t_2) + D$ computed over 10 wave periods. The ratio stays close to 1 for all the simulation cases with some variations.  
    }
    \label{fig:cFp_vs_dEdt}
\end{figure}

The direct wave growth and surface stress extracted from the simulation and introduced in \S\ref{sec:qualitative} offer two ways of computing the energy input rate into the wave $S_{in}$. First, we compute $dE/dt$ from figure \ref{fig:growth_compiled} and correct for the dissipation (\ref{eqn:intro2}); and second we extract the surface pressure $p_s$ and compute the correlation (\ref{eqn:intro1}). 

Figure \ref{fig:cFp_vs_dEdt}(a) shows a comparison of the results obtained using the two methods. The wind input rate $S_{in}(t)$ computed with (\ref{eqn:intro2}) is plotted with dotted lines, and the pressure input rate $S_{p}(t)$ computed with (\ref{eqn:intro1}) is with crosses, for $c/u_* = 2$ and $4$. In both cases, the pressure input $S_p$ closely follows the wave energy growth rate $S_{in}$, although there is a small gap for the $c/u_* = 2$ case. A further demonstration of the dominant role of the pressure term is shown in figure \ref{fig:cFp_vs_dEdt}(b), where we plot the ratio $S_p/S_{in}$ averaged over 10 wave periods for all the cases. The ratio is very close to 1 for most cases, indicating that the pressure input $S_p$ is the major energy input term in $S_{in}$. Again, the smallest wave age cases ($c/u_*=2$) present the largest difference ($S_p/S_{in}=0.8$) and indicate that the wave coherent viscous stress might start to play a role in the strongly forced cases. 

Note that at high $c/u_*$, uncertainties in the budget are related to uncertainties of the decay rate for finite amplitude waves, which get amplified by the large $E$ for the fast travelling waves, together with the very small decay rate which are also hard to accurately capture numerically. Furthermore, the viscous stress input $S_{s}$ could potentially be negative for these fast travelling cases. 


We want to point out that the dissipation correction is necessary in our cases, as the dissipation is non-negligible due to the limited $Re_w$. Although the wave Reynolds number $Re_w$ is constant (and therefore $\gamma_d$ is the same for different cases by (\ref{eqn:gamma_d})), we still have different values of $D$ for different cases of different wave frequency $\omega$ and initial energy $E_0$. The faster travelling waves have higher $E_0$ and therefore higher $D$, and the relative change in energy is much smaller. This relative change in energy (per radian) is reflected by the parameter $\gamma$ (defined in (\ref{eqn:Miles_dEdt})). The underlying assumption of (\ref{eqn:Miles_dEdt}) is that the wave growth is exponential, and $\gamma$ represents the exponential growth rate per radian. In our simulations, we find that the growth rates are so small that for most cases, this exponential growth cannot be distinguished from a linear growth, and the growth rate computed by 
$\gamma^{\prime} = S_{in}/(\omega E_0)$ shows more directly the trend of $S_{in}$. There is an uptake of $S_{in}$ as the instantaneous amplitude slowly increase over the interval of about 10 wave periods for the $c/u_* = 2$ case; in contrast, $S_{in}$ stays almost constant for the  $c/u_* = 4$ case, as the amplitude growth is so small that its effect on $S_{in}$ is negligible.

Overall, we are able to show directly that the pressure energy input plays a dominant role in wave growth for gravity waves of  realistic wave age, especially when there is a finite amplitude established ($ak\geq 0.1$). This is a different picture in term of forcing mechanism from our previous 2D study \citep{WU2021}, where the waves are gravity-capillary waves with $ak=0.05$, and the laminar wind has a linear velocity profile with much stronger shearing ($c/u_*$ around 1).

\subsection{Transient effect and micro-breaking of the strongly forced cases}\label{sec:Fp_evolution}

As we have already seen in figure \ref{fig:growth_compiled} and \ref{fig:cFp_vs_dEdt}, both the wave amplitude and the related wave averaged quantities ($F_p$, $S_{p}$ etc.) are not stationary, especially for the small $c/u_*$ cases. Before we discuss the scaling of these quantities, we show the typical time evolution of wave form drag $F_p$ with two representative cases ($c/u_*=2$ and 8). 

\begin{figure}
    \centering
    \vspace{0.5cm}
    \includegraphics{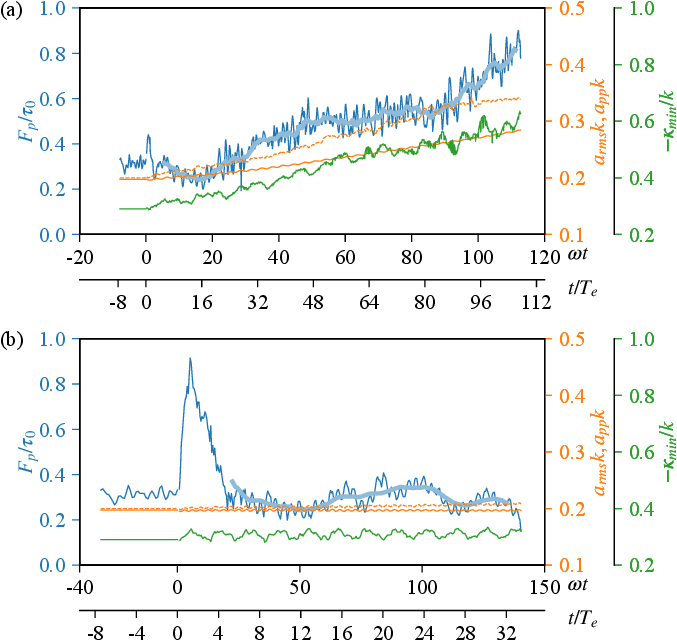}
    \vspace{0.3cm}
    \caption{Time evolution of the wave form drag and wave characteristics, namely steepness and curvature around the crest for (a) $c/u_*=2$ and (b) $c/u_*=8$. The solid orange curves and the dotted orange curves represent the steepness that corresponds to rms wave amplitude $a_{rms}$ and the peak to peak wave amplitude $a_{pp}$ respectively. The green curves represent the curvature around the wave crest $\kappa_{min}$ normalised by wavenumber $k$.}
    \label{fig:Fp_evolution}
\end{figure}

In figure \ref{fig:Fp_evolution} we plot the time evolution of wave form drag $F_p$ (the blue curves) as fraction of the total wind stress, together with a few wave characteristics: the orange curves show the wave amplitude: the solid ones for the rms wave amplitude, defined as $a_{rms} = \sqrt{2}\langle h_w \rangle ^{1/2}$, and the dashed ones for the peak to peak wave amplitude, defined as half of the difference between the peak and the trough $a_{pp} = [max(h_w)-min(h_w)]/2$; the green curves show the curvature around the wave crest, which is taken as the minimum value of the curvature $\kappa_{min}$. These quantities are sampled at  a higher frequency than that shown in figure \ref{fig:growth_compiled} and \ref{fig:cFp_vs_dEdt}.

The $t<0$ part of the curves are from the turbulence precursor where the waves are artificially kept stationary as described in \S\ref{sec:numerics}. After the waves are released, there is a transitory phase where the wave form drag $F_p$ jumps up, but it soon falls back and reaches a stationary level, not far from the precursor one. As we have mentioned in \S\ref{sec:numerics}, we find that the transitory period lasts about 4$T_e$ regardless of the wave frequency $\omega$, which corresponds to the flow adjusting to the initial conditions. Consequently, we do not consider the data for $t/T_e<4$ in our analysis. The ratio of time scale $\omega T_e$ is different for different $c/u_*$, which is why the extent of the transitory period looks different. In general, $T_e$ is much smaller than the wave period.

After $4T_e$, in both cases the wave form drag $F_p$ value fluctuates due to the presence of the turbulence. What is clearly different is that in the $c/u_*=8$ case, the mean value is relatively stable while in the $c/u_*=2$ case, the wave form drag $F_p$ value keeps increasing. The significant increase in $F_p$ is related to the relatively fast wave growth, associated with an increase in the rms amplitude, as well as an increase in the non-dimensional curvature. The curvature $\kappa_{min}/k$ is taken as a measurement of how `sharp' the wave crest is (but not carry information on how three-dimensional the flow is). For the slowest wave case, it significantly increases above the value of the initial third order Stokes wave and later saturates, as shown in \ref{fig:Fp_evolution}(a). The curvature metric could be as important when determining the occurrence of airflow separation (see more in \S\ref{sec:separation}).  Around $\omega t = 80$, the underwater drift transits into turbulence, and the surface develops more prominent 3D structure (ripples), which could also affect the wave form drag.

\begin{figure}
    \centering
    \vspace{0.5cm}
    \includegraphics{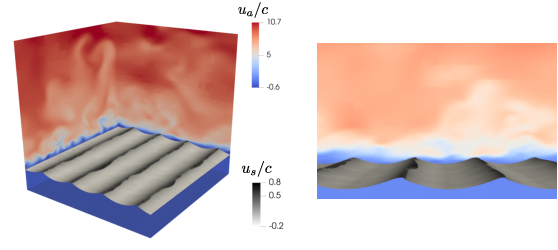}
    \caption{Micro-breaking event around $\omega t=113$. A close-up view shows the micro-breaking features. Initial $ak=0.2$, $c/u_*=2$. The instantaneous rms steepness $a_{rms}k=0.285$, and the instantaneous peak to peak steepness $a_{pp}k=0.339$.}
    \label{fig:micro-breaking}
\end{figure}

At later time (after  around $\omega t = 90$), the rms amplitude is stills increasing even though the peak to peak amplitude starts to plateau. The saturation is due to the micro-breaking of the waves, which is shown in figure \ref{fig:micro-breaking}. This micro-breaking behaviour is characterised by a confined collapse of the water surface near the crest. This coincides with a sharp increase in the wave drag $F_p$. We have only ran one case for long enough time to observe the whole history of wave growing till the point of breaking. In appendix \ref{sec:appendix0_1}, we include another case with initial $ak=0.3$ and $c/u_*=2$, which exhibits a quite different behaviour, in terms of associated form drag. It does not take too long before reaching the breaking point, and the breaking is much more perceivable (spilling breakers) with droplets ejection. There is a reduction of wave form drag $F_p$ instead of an increase because of the sharp decrease of wave amplitude. A systematic study of the effect of micro-breaking on the form drag within this fully coupled approach is left for future studies.

This highlights the importance of a fully coupled approach, especially for the strongly forced condition. For the discussion in \S\ref{sec:pressure_distribution}, however, we focus on the effect of the initial condition $ak$ and $c/u_*$ on the surface pressure distribution. This is done by taking a small enough averaging window after $t/T_e>4$ so that the transient effect is not prominent, and $a_{rms}(t)k$ is close enough to $ak$. 
For example for the case shown in figure \ref{fig:Fp_evolution}(a) we take the window of time $\omega t \in [10,30]$, and for the case shown in figure \ref{fig:Fp_evolution}(b) we take the window of time $\omega t \in [25,130]$. Our results can then be compared to previous numerical studies where the motion and shape of the waves are prescribed, as well as to the experimental results. In \S\ref{sec:bulk_parameter} when the $F_p$ and $S_{in}$ scalings are concerned, we bring some of the transient effect back into discussion.

\section{Surface pressure distribution} \label{sec:pressure_distribution}
\subsection{Definitions}

To understand better the dynamics of the wind wave interaction and to compare with theoretical formulations introduced in \S\ref{sec:intro_theory}, we proceed by analysing the detailed structure of the surface pressure distribution $p_s$. This is done by extracting the principal mode of the Fourier transformation, a method that was also used in previous numerical work, see e.g. \citet{KIHARA2007a,DRUZHININ2012a}. The structure of the pressure field is shown in figure \ref{fig:pressure_shear}(a) and clearly contains wave-induced signals, while also being influenced by the instantaneous turbulence. To distinguish the wave-induced effect from the turbulent fluctuation, we introduce phase averaging. For any quantity $q(x,y,z)$, the phase average is
\begin{equation} \label{eqn:bar_q}
    \bar{q}(\theta,z) = \frac{1}{N_{\text{w}} L_0}\sum_{n=1}^{N_{\text{w}}-1} \int_{-L_0/2}^{L_0/2}dy \; q(x = \lambda(n+\theta/2\pi),y,z),
\end{equation}
where $\lambda = 2\pi/k = L_0/4$ is the wavelength of the initial waves, and $N_{\text{w}}=4$ is the number of waves in the $x$ direction. The phase $\theta$ can be extracted from the surface elevation $h_w(x,y,t)$ and is therefore generalizable to cases which are not strictly sinusoidal.

The surface pressure can be generally described as the sum of different frequency modes,
\begin{equation} \label{eqn:my_p}
	p_s(\theta, t) = \sum_{n=1}^{\infty} \hat{p}_n\cos(n\theta + \phi_{pn})
\end{equation}
where $\phi_{pn}$ is the pressure phase shift and $\hat{p}_{n}$ is the pressure amplitude of mode $n$. Meanwhile, the surface elevation can be written as
\begin{equation}
	h_w (\theta, t) = \sum_{n=1}^{\infty} a_n \cos (n\theta +\phi_n),
\end{equation}
with $h_w (\theta, t) \approx a\cos (\theta)$ since the surface elevation $h_w$ is largely monochromatic in our simulation (and we can always shift the reference point so that the phase $\phi_1$ is zero). 

Once given the surface pressure distribution $p_s$ (\ref{eqn:my_p}), the wave form drag  $F_p$ (\ref{eqn:momentum_partition}) becomes
\begin{align}\label{eqn:Fp_sin}
    F_p = \langle p_s\frac{\partial h_w}{\partial x} \rangle 
    &\approx \sum_{n=1}^{\infty} \hat{p}_{n} \:a_n nk \langle \cos(n\theta+\phi_{pn})\sin(n\theta + \phi_n) \rangle \\
    &= \hat{p}_{1} \:ak \langle \cos(\theta+\phi_{p1})\sin(\theta) \rangle = \frac{1}{2}\:ak\:\hat{p}_1\sin(\phi_{p1}),
\end{align}
and $S_p$ follows as $S_p=cF_p$.
Finding the drag force and the pressure input rate now simplifies to finding the pressure perturbation amplitude $\hat{p_1}$ and the phase shift $\phi_{p1}$ that correspond to wave number $k$. Notice how a non-zero phase shift $\phi_p$ is necessary for a non-zero $F_p$ and $S_p$.
Since (\ref{eqn:Fp_sin}) shows that only the principal mode ($n=1$) contributes to the wave growth, we then focus on how $\hat{p_1}$ and $\phi_{p1}$ depend on $c/u_*$ and $ak$ qualitatively.

\subsection{Streamline and asymmetric pressure patterns}

\begin{figure}
    \centering
    \includegraphics{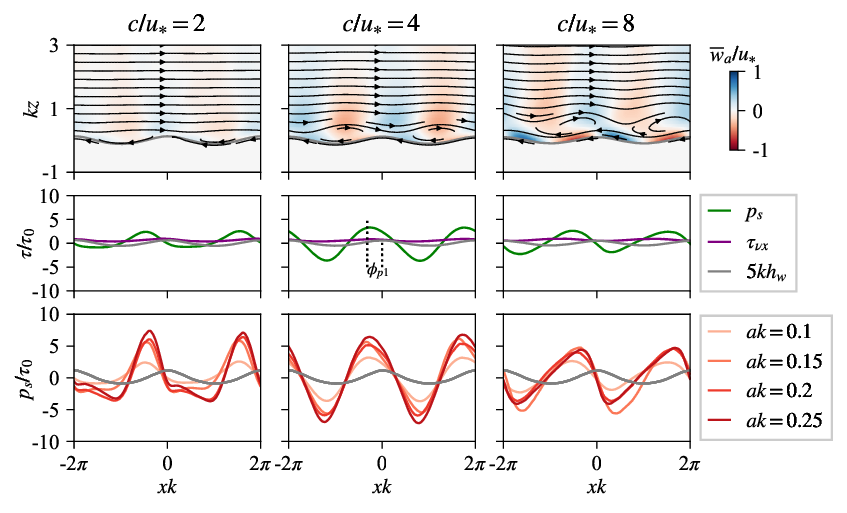}
    \caption{Vertical velocity field, streamline and 1D stress distribution for three different wave ages $c/u_*=2,4,8$. Top row (initial $ak=0.1$): solid black lines in the top row are streamlines in the moving wave frame of reference (i.e. plotted with $\bar{w}$ and $\bar{u}-c$), and the colour shows the phase averaged vertical velocity $\bar{w}$. Notice that the higher the $c/u_*$, the further the wave induced perturbation extends above the waves. Middle row (initial $ak=0.1$): The asymmetric pressure distributions (green lines) that result from the distorted streamlines. The purple line is the shear stress. The phase shift $\phi_p$ between the pressure $p_s$ and the water surface elevation $h_w$ gives rise to the drag force and energy input. 
 Bottom row: the shape of the $p_s$ distribution is consistent across different steepness, shown by different colours. The amplitude, however, seems to increase from low ($ak=0.1$) to moderate ($ak=0.15$) initial steepness, but not change much from moderate to high initial steepness ($ak=0.2,0.25$). The grey lines in all plots indicate the wave surface position, with exaggerated steepness.}
    \label{fig:streamline}
\end{figure}


Figure \ref{fig:streamline} top row shows the phase averaged vertical velocity $\bar{w}$, for three flow conditions ($c/u_*$=2,4,8; $ak$=0.1). The alternating patterns demonstrate the perturbation by the waves, as opposed to uniform zero for a flat surface. In the slowest wave cases (i.e. $c/u_* =2$), the alternating sign mostly comes from the straining and relaxing of the shear layer (because the airflow follows the boundary shape). In the intermediate wave speed cases ($c/u_* =$ 4 and 8), the wave orbital velocity becomes significant and it leaves an imprint on the airflow (because the airflow follows the vertical motion of the boundary). Here we are plotting below $kz=3$, however we noticed that the wave induced perturbation in $\bar{w}$ extends higher up with increasing $c/u_*$, to almost $kz=2\pi$, in the $c/u_*=8$ case.

In figure \ref{fig:streamline} top row we also plot the streamlines in the wave frame of reference (i.e with $\bar{w}$ and $\bar{u} - c$). There are recirculation cells because the vertical velocity is of alternating signs, and the horizontal velocity changes sign at some height. This height is often called the critical height, and it depends on the value of $c/u_*$. The higher $c/u_*$ is, the further away the critical height is from the water surface. These recirculating cells influence the pressure variation $p_s$ at the water surface in a complicated way, which is plotted in the middle row of figure \ref{fig:streamline} with green lines. 

The middle row of figure \ref{fig:streamline} is the averaged stress distribution for both the pressure and the viscous shear stress (shown in fig. \ref{fig:pressure_shear}). We see clearly that the pressure maximum is on the windward face, and the phase shift is marked by $\phi_{p1}$. Notice that even for the smallest steepness case ($ak=0.1$), the shapes of the pressure distribution are not sinusoidal.
For example, at $c/u_*=2$, the trough of the pressure signal is rather flat, which is a sign of sheltering (with or without a certain level of airflow separation). For the $c/u_*=8$ cases, the pressure distribution is tilted forward. Only in the $c/u_* = 4$ case does the pressure distribution roughly resembles a sinusoidal wave.
The pressure structures are in qualitative agreement with those found in simulations \citep[][]{YANG2010,KIHARA2007a} and experiments \citep[][]{MASTENBROEK1996} with corresponding $c/u_*$. The non-sinusoidal pressure shape is the signature of higher frequency modes and would contribute to the growth of corresponding wave frequencies. 

The bottom row shows how the 1D pressure distribution changes with different initial $ak$, ranging from $0.1$ to $0.25$ (colour coded). The shapes are similar for the same $c/u_*$, with the amplitude of the pressure variation increasing with wave steepness $ak$. The largest difference is between $ak=0.1$ and the other three $ak$ values, where the amplitude of the pressure seems to saturate at high $ak$.

\subsection{Pressure amplitude and phase shift} \label{sec:pressure_amp_phase}

\begin{figure}
    \centering
    \includegraphics{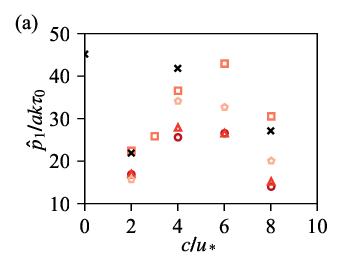}
    \includegraphics{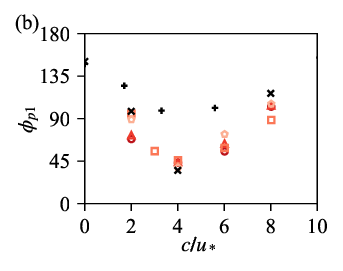}
    \includegraphics{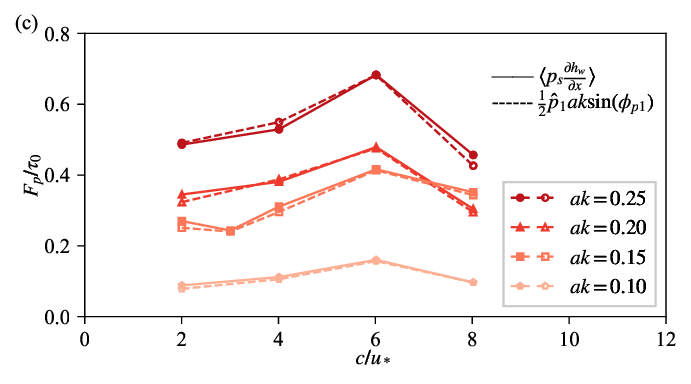}
    \caption{(a): Pressure amplitude $\hat{p}_1$ normalised by the nominal wall stress $\tau_0 = \rho_a u_*^2$, and in addition $ak$, plotted against $c/u_*$. (b): Pressure phase shift $\phi_{p1}$ as a function of $c/u_*$. Notice that because of the $F_p=(1/2)\hat{p}_1\:ak\sin(\phi_{p1})$ relation, the drag force is the largest when $\phi_p = 90^{\circ}$, and zero when $\phi_p = 0^{\circ}$ or $180^{\circ}$. The results from \citet[][]{KIHARA2007a} of $Re_{\lambda} = 161$ and $ak=0.1$ are plotted with black crosses. The results from \citet[][]{DRUZHININ2012a} of $Re = 10000$ and $ak=0.2$ are plotted with black plus signs. (c)The wave form drag $F_p$ is not a strong function of $c/u_*$ for all values of the steepness $ak$. We also show the full integral value $\langle p_s \partial{h_w}/\partial{x}\rangle$ in comparison to the single mode representation $(1/2)\hat{p}_1ak\sin(\phi_{p1})$. The markers and colours are the same with those in figure \ref{fig:cFp_vs_dEdt}(b) and \ref{fig:streamline}.
    }
    \label{fig:phat_dphase}
\end{figure}

Figure \ref{fig:phat_dphase} shows the pressure amplitude $\hat{p}_1$ and phase shift $\phi_{p1}$ as a function of $c/u_*$ and $ak$. 
These quantities are computed by Fourier transform of the phase averaged surface pressure $p_s$. The `surface' is defined as the wave following surface $4\Delta = 0.1/k$ away from the air water interface.
We have tested the sensitivity to the location within the first 8 grid points and it does not present much difference (as long as we are in the viscous layer). 


Figure \ref{fig:phat_dphase}(a) shows that the amplitude $\hat{p}_1$ first increases with $c/u_*$ until $c/u_*\approx 6$ and then decreases, for all steepness $ak$. Figure \ref{fig:phat_dphase}(b) shows that the phase shift $\phi_{p1}$ follows the opposite trend. The net result is that the the drag force shown in figure \ref{fig:phat_dphase}(c) is not a strong function of $c/u_*$, which is in agreement with previous studies in the slow wave regime. Figure \ref{fig:phat_dphase}(c) also confirms (\ref{eqn:Fp_sin}): the dotted and solid lines show the single mode representation and the integral representation of wave form drag $F_p$ respectively, which agree very well, even when the pressure distribution is not necessarily sinusoidal.

Taking a closer look at the phase shift, it is around 90 degree for the strongest forcing cases $c/u_*=2$, and then goes under 90 degree between $c/u_*=2$ and $6$, and then slightly above 90 degree at $c/u_*=8$. This indicates that the sheltering mechanism is dominant in the strong forcing conditions, and that the theories based on linear stability analysis might be at work in the higher wave age cases (more on this in \S\ref{sec:discussion_theory}). Good agreement was found with results from \citet[][]{KIHARA2007a} (marked with black crosses) at $ak=0.1$. 
The configuration in \citet[][]{KIHARA2007a} is similarly a pressure-driven channel flow, with $Re_{\lambda}=161$. 
Data from \citet[][]{DRUZHININ2012a} show larger $\phi_{p1}$ for all wave age although the trend with wave age is similar. They used a bulk Reynolds number of $10,000$ with a Couette flow configuration. We could not infer the exact friction Reynolds $Re_{\lambda}$ value from the information provided in the paper, but they should be of the same order of magnitude as in our setup.

The pressure amplitude $\hat{p}_1$ is normalised by $ak\tau_0$ in figure \ref{fig:phat_dphase}(a), and this choice is made by a commonly adopted scaling argument. Intuitively, and also used in the theoretical studies mentioned in \S \ref{sec:intro_theory}, the pressure variation amplitude should scale with $\rho_a (ak) U_{ref}^2$, with $U_{ref}$ being some characteristic wind velocity (not necessarily the friction velocity $u_*$). From figure \ref{fig:phat_dphase}(a) we see that this scaling does not collapse $\hat{p}_1$ with respect to $ak$, at least not when $u_*$ is used. Now defining the ratio between $\hat{p}_1$ and $ak \rho_a u_*^2$ as $P$,
\begin{equation}\label{eqn:ratioP}
P = \hat{p}_1/ak \rho_a u_*^2.
\end{equation} 
This ratio $P$ represents $(U_{ref}/u_*)^2$, the ratio between the should-be characteristic velocity $U_{ref}$ and the friction velocity $u_*$. From figure \ref{fig:phat_dphase} (a) we see that $P$ ranges from around 15 to 45, indicating that $U_{ref}/u_*$ is around 4 to 7. 

\subsection{A note on airflow separation and micro-breaking for steep waves} \label{sec:separation}
\begin{figure}
    \centering
    \includegraphics{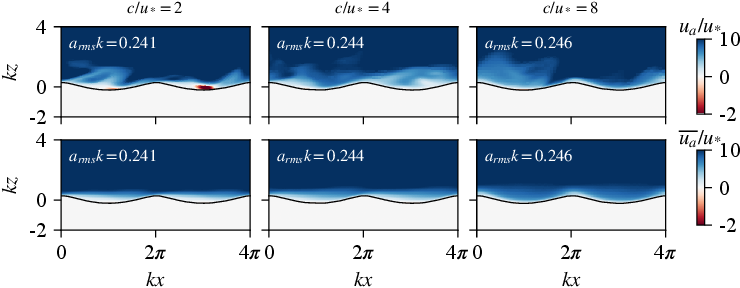}
    \caption{The instantaneous horizontal velocity (first row) at $y=0$ and the phase averaged horizontal velocity (second row) in laboratory frame of reference, for three different wave ages at comparable instantaneous steepness $a_{rms}k$. There is airflow separation for the $c/u_*$ case but the effect is intermittent localised. For the $c/u_*=4$ and $8$ cases there is no separation at similar steepness. The second row shows that the phase averaged flow field $\overline{u}_a$ is similar in effect to that of a non-separated case.}
    \label{fig:separation}
\end{figure}

We observe intermittent airflow separation when the wave steepness $a_{rms}k$ reaches a value between around 0.23 to 0.27, for the $c/u_*=2$ and 4 cases. In the first row of figure \ref{fig:separation}, we show examples of the instantaneous horizontal velocity at the center slice ($y=0$), for three wave ages $c/u_*=2,4,8$, when the $a_{rms}k$ steepness is around 0.24. There is airflow separation and recirculation for the $c/u_*=2$ case, indicated by the confined negative $u$ zone. In contrast, there is no such airflow separation for the $c/u_*=4$ and 8 case (or much rarer throughout the whole domain), suggesting that the increased regular wave induced motion (which scales with $akc$) near the bottom boundary could suppress the airflow separation. The airflow separation in the $c/u_*=2$ case is, however, highly intermittent and localised. In fact, for the phase averaged horizontal velocity shown in the second row, the airflow separation is not distinguishable.

Notice that the micro-breaking mentioned in \S\ref{sec:Fp_evolution} does not occur until $a_{rms}k \approx 0.3$. This means that airflow separation can occur before the waves break, due to the sharp directional change of the lower boundary, when the waves are steep and short-crested. This finding is consistent with the previous experimental and numerical findings \citep[]{BUCKLEY2020a,SULLIVAN2018b,DRUZHININ2012a,YANG2010,DONELAN2006}.
We comment that it is, however, not practical to determine an exact steepness value of $a_{rms}k$ at which separation starts to occur. It is likely also dependent on other geometric quantities such as $\kappa_{min}/k$, as they are a more local measure of the change of direction in the boundary, and therefore closely related to vorticity generation in the boundary layer \citep{BATCHELOR2000}. In fact in \citet[]{BUCKLEY2020a} figure 16, the \emph{likelihood} of airflow separation is reported experimentally, and it increases with steepness but decreases with wave age, which is what we observe as well. We caution that the occurrence of separation and the exact separation point can also depend on the Reynolds number of the flow, which is much lower in the DNS than in realistic wind wave airflow.

Nonetheless, the onset of airflow separation does not significantly affect the discussion in \ref{sec:pressure_amp_phase}. In fact, if we consider the phase averaged velocity and surface pressure, the separated and non-separated sheltering cases exhibit similar features. That is to say, even the separating cases can be readily incorporated into the current framework of principal $p_s$ mode analysis. Although it is possible that separation point might shift the phase $\phi_{p1}$, and this explains why for $c/u_*=2$, the $ak=0.2$ and $0.25$ cases have different $\phi_{p1}$ from the $ak=0.15$ and $0.1$ case (see figure \ref{fig:phat_dphase}(b)). 

\section{Scaling the wave form drag $F_p$ and the energy input rate $S_{in}$} \label{sec:bulk_parameter}
In this section, we discuss the scaling of the wave form drag $F_p$ and the energy input rate $S_{in}$ as functions of $c/u_*$ and $ak$, and compare our results to those from the literature. Apart from the initial steepness $ak$, we also discuss the time dependent $a_{rms}k$, and especially its effect on the wave form drag $F_p$.

\subsection{Wave drag $F_p/\tau_0$} \label{sec:discussion_drag}
\begin{figure}
    \centering
    \includegraphics{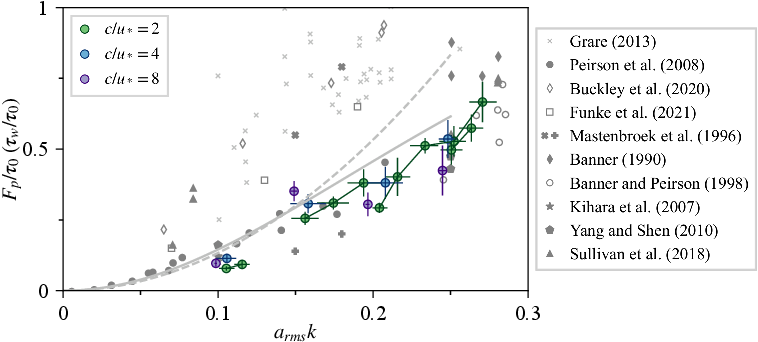}
    \caption{The wave form drag $F_p$ (or in some cited works wave drag $\tau_w$ defined by (\ref{eqn:tau_w})) as fraction of the total stress $\tau_0$, plotted as a function of rms steepness $a_{rms}k$. For the $c/u_*=2$ cases (green points), we take multiple averaging windows because of the transient evolution of $F_p$. The bar in $x$ axis is the range of $a_{rms}k$ in the averaging time window. The bar in $y$ is the standard deviation of $F_p$ fluctuation (mostly due to turbulent fluctuation). Points that belong to the same initial $ak$ case are connected with a line.
Other numerical data: stars \citet[][]{KIHARA2007a}, $c/u_*=2,4,8$, mostly overlapping with the $ak=0.1$ results; pentagons, \citet[][]{YANG2010}, $c/u_*=2$. Experimental results: solid circles, \citet[][]{PEIRSON2008}; solid crosses, experimental observation from \citet{MASTENBROEK1996}; plus signs, numerical prediction from \citet{MASTENBROEK1996}; solid diamonds, \citet{BANNER1990}; open circles, \citet{BANNER1998}; light crosses, \citet[][]{GRARE2013a}; open diamond, \citet[][]{BUCKLEY2020a}; open squares, \citet[][]{FUNKE2021}. The last three data sets denoted with open marks are purely wind generated waves, and the \citet[][]{GRARE2013a} data set has mixed types, while the others are all mechanically generated waves (or similar numerical setups). The \citet{BANNER1990} and the \citet{BANNER1998} datasets include waves with micro-breaking. Dashed line: the quadratic representation $F_p = 1/2\beta(a_{rms}k)^2$ with a constant $\beta$; solid line: the Belcher correction (\ref{eqn:beta_correction}).}
    \label{fig:drag_coefficient}
\end{figure}

In \S\ref{sec:pressure_amp_phase}, we have shown that the drag is not a strong function of $c/u_*$ in the slow wave regime. However, it is strongly dependent on the steepness. Instead of showing the wave form drag $F_p$ as a function of initial $ak$, figure \ref{fig:drag_coefficient} shows the drag coefficient $F_p/\tau_0$ as a function of the rms steepness $a_{rms}k$. Since for the small wave age cases (the green dots), there is a significant increase of $a_{rms}k$ due to the wind forcing, we take multiple averaging windows. The points that belong to the same case are connected with a line. The bar in $x$ axis is the range of $a_{rms}k$ in the averaging time window.  The bar in $y$ axis is the standard deviation of $F_p$ fluctuation, which is mostly due to the turbulent fluctuation (see figure \ref{fig:Fp_evolution}).

Again if we analyse the first data point of each green dots group and the blue and purple dots of the same initial $ak$, they are close to each other, as we have found in \S\ref{sec:pressure_amp_phase}, meaning that the the ratio $c/u*$ has little effect on the wave form drag. For the small steepness regime ($ak<0.2$), the data roughly scales with $(ak)^2$, with some small variation in different $c/u_*$,
\begin{equation}
    F_p \sim (ak)^2 \tau_0.
\end{equation}
More specifically the prefactor is $1/2P\sin(\phi_p)$, with $P$ defined in (\ref{eqn:ratioP}).
For higher steepness $ak=0.2, 0.25$, we see a plateau in $F_p/\tau_0$ and a departure from the $(ak)^2$ scaling, and slightly larger variation with $c/u_*$. 


However, if we trace an individual case of $c/u_*=0.2$, the picture is further complicated. For example, for the $ak=0.2$ case, the $F_p$ value undergoes stages of growth and increases from $0.25\tau_0$ to around $0.7\tau_0$ over the course of $a_{rms}k$ from 0.2 to 0.27. The time evolution of the strongly forced cases overall seems to better fits the $(a_{rms}k)^2$ scaling than the ensemble of cases, which falls shorts of the $(ak)^2$ scaling. There also seems to be a wave history effect: for example, the initial $ak=0.15$ case shows higher value of $F_p/\tau$ when $a_{rms}(t)k$ reaches 0.2, when compared to the case with initial $ak=0.2$. This is probably related to the wave geometry and its short-crestness that evolves with the amplitude growth, that is not captured by only varying the initial amplitude for the Stokes waves.

Figure \ref{fig:drag_coefficient} also shows numerical and experimental data from the literature. If only considering the initial $ak$ effect, our results agree very well with previous numerical studies across different $ak$. The $ak=0.1$ results is very close to those from \citet[][]{KIHARA2007a}, and the $ak=0.25$ results are within the range of those reported by \citet[][]{YANG2010}. Note that these simulations are performed with prescribed wave boundary shape and motion. This agreement serves as a further validation for the current numerical method. On the other hand, it suggests that the one-way coupled approach could suffice for predicting the wave form drag of weakly coupled cases where the waves' growth is very slow. The necessity of the fully coupled approach comes when the waves are strongly forced, grow relatively fast and exhibits strong nonlinear behaviour such as short-crestness and micro-breaking.

For comparison with experimental studies, we note that some of the data plotted in figure \ref{fig:drag_coefficient} are actually $\tau_w$ defined by (\ref{eqn:tau_w}) instead of $F_p$. Since we have already verified that the pressure is responsible for over 80\% of the energy flux, the $F_p$ and $\tau_w$ values do not differ by much for the cases discussed here; at least the small difference does not affect the general trend of $F_p/\tau_0$ with increasing $ak$. 

\citet[][]{PEIRSON2008} (solid circles) measured $\tau_w$ by the spatial wave energy growth, and their data match with ours quite well. They also suggested a correction to the $(ak)^2$ relation inspired by \citep[][]{BELCHER1999}, with two fitted parameters $\beta_f$ and $\beta_t$
\begin{equation} \label{eqn:beta_correction}
    \tau_w/\tau_0 = (\beta_f+\beta_t)(ak)^2/[2+\beta_f (ak)^2]
\end{equation}
that seem to fit a compilation of the data sets well (see their figure 5). This correction is plotted with the solid line in figure \ref{fig:drag_coefficient}. 
The other experimental studies have reported a wave drag coefficient somewhat higher. \citet[][]{MASTENBROEK1996} measured the wave drag coefficient by using a fixed pressure probe at a fixed height $kh=\pi$, and \citet[][]{GRARE2013a} used PIV viscous stress measurement, pressure with fixed or wave following probe for different subset data. \citet[][]{BUCKLEY2020a} and \citet[][]{FUNKE2021} were obtained from the same data set; \citet[][]{BUCKLEY2020a} used PIV viscous stress measurement and computed pressure as a stress residual, while \citet[][]{FUNKE2021} reconstructed the pressure field by solving the Poisson equation. 

From the synthesis of data, we can see that the numerical estimations of $F_p/\tau_0$ are in general lower than the experimental measured ones. Since we have seen that there is a wave history effect which is related to the evolving wave shape, it explains why numerical simulations with more idealized wave shape (Airy waves or Stokes waves) might be missing that effect and therefore predicting lower wave form drag $F_p$. \cite{YANG2010} noticed that nonlinearity can play an appreciable effect by comparing their Airy waves and Stokes waves results. The current work further shows that the steep wave shape can deviate even more from the Stokes waves and increase the wave form drag. However, there remains significant scatters within the experimental data using different methods to measure the stress. The ones inferred from the wave growth seem to be consistently lower than the ones measured from the air stress in experiments, and the differences are beyond the scatters that might be introduced by different wave ages. Although we have verified using our simulation that the measurement of $F_p$ directly from the air stress or indirectly from the wave growth should be consistent, there remain a few possible reasons for the scatters in the experimental data: one is the existence of 3D smaller scale waves (roughness elements) that increases the drag; the other is the uncertainty caused by the air side measurement, especially the pressure extrapolation error from a finite height to the surface, discussed in \citet[][]{DONELAN2006} and \citet[][]{GRARE2013a}. A further examination of the extrapolation error will require a study of the vertical pressure structure.



\subsection{Growth rate $\gamma$}
\begin{figure}
    \centering
    \includegraphics{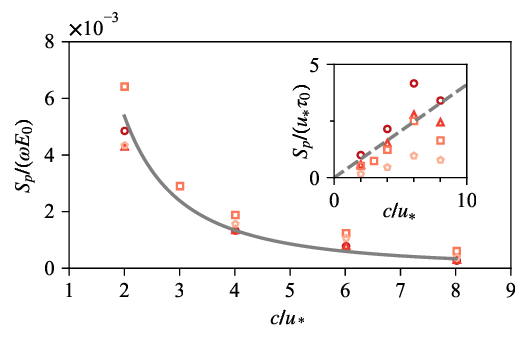}
    \caption{Non-dimensional growth rate scaling computed with the points in figure \ref{fig:phat_dphase}. Inset figure shows the energy input rate $S_p$ increasing with increasing $c$, while the main plot shows the non-dimensional growth rate ($\gamma$) decreasing with increasing $c/u_*$. The grey line is the $(u_*/c)^2$ fitting. It demonstrates that the non-dimensional growth rate scaling is dominated by the $\omega E$ normalisation.}
    \label{fig:energy_scaling}
\end{figure}

The energy input by pressure is closely linked to the wave form drag by $S_p = cF_p$; or considering the more general definition of wave drag (\ref{eqn:tau_w}), the total wind input is $S_{in} = c\tau_w$. The two are used interchangeably in the present discussion, i.e. $S_{in} = c\tau_w \approx cF_p$.

We have seen that the drag force $F_p$ is not a strong function of $c/u_*$, so the pressure energy input rate $S_{p} = cF_p$ increases with $c/u_*$ as shown in the inset of figure \ref{fig:energy_scaling}, i.e. in the slow wave regime, the energy flux is higher for waves travelling faster (at a fixed $u_*$). This could appear in contradiction to the observation that the slowest travelling waves have the fastest growing energy curve in figure \ref{fig:growth_compiled}. This is however not self-contradicting, because the curves in figure \ref{fig:growth_compiled} reflect the \textit{relative} rate of change of energy, which is $S_{in}$ further normalised by the total energy $E$, and $E$ is larger for faster waves. (Note that in figure \ref{fig:growth_compiled}, since we consider the net energy growth, another factor that is the viscous decay is also larger for the faster waves, since $\gamma_d$ is constant in our simulation). 



This normalisation by the total energy $E$ and angular frequency $\omega$, i.e. the definition of growth rate per radian $\gamma = S_{in}/(\omega E)$, was introduced by \citet[][]{MILES1957}, and is based on the assumption that the growth is exponential. Considering the definitions of wave energy and the gravity wave dispersion relation,
\begin{equation}
    E = \frac{1}{2}\rho_w g a^2, \;\; \omega = kc = \sqrt{gk}
\end{equation}
and using the assumption that $F_p \sim (ak)^2$ (which we have seen to be questionable at high $ak$), and by introducing the prefactor $\beta$ \citep[][]{MILES1957}, we obtain
\begin{equation} \label{eqn:Fp_beta}
    F_p = \frac{1}{2} \beta (ak)^2 \tau_0 = \frac{1}{2} \beta (ak)^2 \rho_a u_*^2,
\end{equation}
which becomes
\begin{equation}  \label{eqn:Sin_scaling}  
    \gamma = \frac{S_{in}}{\omega E} = \frac{cF_p}{\omega E} = \beta \frac{\rho_a}{\rho_w}\left(\frac{u_*}{c}\right)^2.
\end{equation}
It is worth noticing that this relationship, widely used in the literature, presents some strong self-correlation between the normalisation of $S_{in}$ by $\omega$ in the left hand side and the phase speed $c=\omega/k$ on the right hand side. The resulting $(u_*/c)^2$ scaling is reflected in figure \ref{fig:energy_scaling}. 

\begin{figure}
    \centering
    \includegraphics{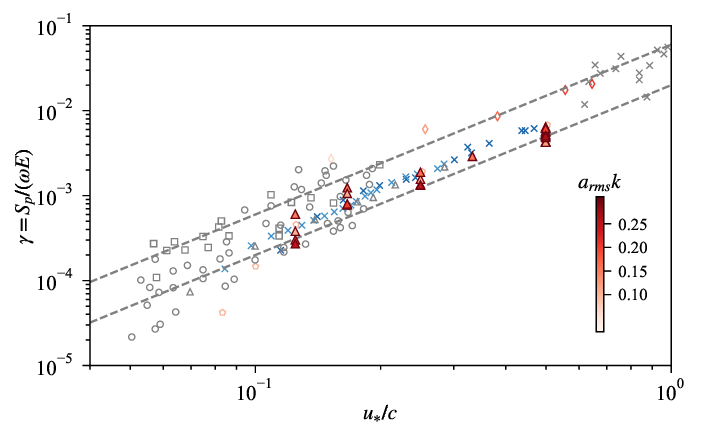}
    \caption{Growth rate parameter $\gamma$ as function of inverse wave age $u_*/c$. The value of $a_{rms}k$ is denoted with the color scale. Notice that the added averaging windows in figure \ref{fig:drag_coefficient} result in more points for the $c/u_*=2$ cases, but the $\gamma$ values are very close to each other, due to the fact that the time evolving $F_p(t)$ scales with $(a_{rms}(t)k)^2$ relatively well. The points from cited works are also colour-coded whenever the steepness value can be identified. Numerical works: blue crosses, \citet[][]{YANG2013} with JONSWAP spectrum; open triangles, \citet[][]{KIHARA2007a}, $ak=0.1$. Experimental works: open diamonds, \citet[][]{BUCKLEY2020a}, $ak$ values as the colours indicate; grey symbols: data compiled by \citet[][]{PLANT1982} with no steepness information. Dotted lines, the range of $\beta$ proposed by \citet[][]{PLANT1982} based on empirical evidence. }
    \label{fig:energy_comparison_literature}
\end{figure}

The representation of (\ref{eqn:Sin_scaling}) in figure \ref{fig:energy_comparison_literature} is often taken as an indirect proof of  Miles' theory. \citet[][]{PLANT1982} compiled laboratory and field measurements known to the date (plotted in grey symbols in figure \ref{fig:energy_comparison_literature}), which became the benchmark and established the $(u_*/c)^2$ scaling, although the empirical range of $\beta$ (indicated in grey dotted lines) is higher than the original prediction from \citet[][]{MILES1957}. 

We caution that while the $(u_*/c)^2$ scaling seems to hold, there is a wide scatter in the $\beta$ value at a given value of $u_*/c$, with sometimes over an order of magnitude variation. We also note that alternatives for the reference velocity have been proposed (e.g. the sheltering coefficient at half wavelength by \citet[][]{DONELAN2006} or the middle layer velocity from \citet{BELCHER1999}), and the reported values of the $\beta$ parameter by experimental and numerical studies could be presented in terms of another reference velocity, leading to estimations of the sheltering coefficient (see \citet{PEIRSON2008} and \citet[][]{YANG2013} for example).


A large contributing factor to the scatter is the role of the wave steepness at a given wave age, as already discussed by \citet[][]{PEIRSON2008, BUCKLEY2020a}. The steepness is indicated in figure \ref{fig:energy_comparison_literature} with different shades of red for the data sets where the wave steepness can be identified.
As we have mentioned, the assumption that the wave form drag scales with the steepness $(ak)^2$ does not hold for moderate to high steepness ($ak>0.15$).


The other factor is again, the uncertainty in the pressure-slope correlation (\ref{eqn:intro1}) measurements. The data sets compiled by \citet[][]{PLANT1982} were all obtained by measuring the aerodynamic pressure, with either fixed or wave following probes. This is to some extent due to the difficulty in directly measuring the wave growth as an alternative: for the fast moving waves, measuring the extremely small growth in amplitude is prone to errors; and for the less controlled field campaigns, it is hard to single out the wind input from the nonlinear interactions and dissipation. It is of crucial importance, therefore, that we find ways to quantity the uncertainties in these pressure measurements.

In summary, the $(u_*/c)^2$ scaling in figure \ref{fig:energy_comparison_literature}, despite being robust because of the normalisation, inherits the uncertainty reflected in figure \ref{fig:drag_coefficient}. \textcolor{black}{The normalisation of $S_{in}$ by $\omega$ and $E$ following (\ref{eqn:Miles_dEdt}) is questionable with the growth rate being very small due to the small density ratio $\rho_a/\rho_w$ so that the exponential growth cannot be verified in a convincing way; and the normalisation makes the $\gamma$ parameter too skewed by the wave characteristics.}

We want to mention that it remains to be studied how the results from the current study and the other lab experiments with nearly monochromatic wave trains can be extended to broadband ocean waves spectrum. The method to date \citep[][]{SNYDER1981,DONELAN2006,YANG2013} is to keep the linear assumption, and the correlation term (\ref{eqn:intro1}) becomes the cross-spectrum
\begin{equation} \label{eqn:Q}
    Q(\omega) = \langle p_s(\omega)h_w(\omega)* \rangle.
\end{equation}
Interestingly, the numerical study of a broad spectrum wave field from \citet[][]{YANG2013} (blue crosses) reported growth rate of very similar magnitude to our study.
The numerical methods are very different: the points from \citet[][]{YANG2013} are from computing (\ref{eqn:Q}) in one run for different wave frequency $\omega$, while the points in our study are from different runs with different initial $c/u_*$ and $ak$. The steepness $a(\omega)k$ is not reported in \citet[][]{YANG2013}, therefore it is hard to draw a definite conclusion. 





\begin{toexclude}
and $\phi_p = \arctan(\frac{u_*^2}{U_m^2}\beta) + \pi$, and $\hat{p} = \rho_a ak  U_m^2$
Miles theory relates growth to some property at the critical layer, because of the global property of the solution of the boundary-value problem (Rayleigh equation).
\end{toexclude}

\section{Discussion}

\subsection{The range of phase shift $\phi_p$ and implications for potential theories} \label{sec:discussion_theory}
Based on the $\hat{p}_1$ and $\phi_{p1}$ results, we discuss the implication of numerical results for different theories mentioned in the introduction \S\ref{sec:intro_theory}.
The air pressure distribution is of critical importance to understanding both the wave form drag and the wave growth. It also provides insights into the airflow structure, and therefore can be used to validate or invalidate theories. By comparing our real number representation (\ref{eqn:my_p}), (\ref{eqn:ratioP}) to the complex number representation (\ref{eqn:Miles_p}) there is the correspondence that

\begin{equation}
    \beta = P\sin(\phi_{p1}),  \; \alpha = P\cos(\phi_{p1})
\end{equation}
and
\begin{equation}
    \phi_{p1} 
	\begin{cases}
        = \pi/2 & \text{if} \ \alpha=0 \\
        \in (0, \pi/2) & \text{if} \ \alpha/\beta > 0 \\
        \in (\pi/2, \pi)  & \text{if} \ \alpha/\beta < 0 \\  
	\end{cases}
\end{equation}
We have based the discussion around the imaginary part of the pressure distribution $\beta$, which is the $90^{\circ}$ out of phase part with the surface (i.e. in phase with the surface slope). It is always positive for the slow moving waves because of the direction of the energy flux. The real part $\alpha$, although not contributing to the growth, is informative if we want to determine the phase $\phi_p$. 

There has not been much discussion on $\alpha$, although recently \citet[][]{BONFILS2021} used an asymptotic method to solve the Rayleigh equation and they pointed out that the real part $\alpha$, which is often neglected, changes the wave phase speed, and that $\alpha$ can be positive. They have also argued that for strong forcing case, $\alpha$ is around 0, which validates Jeffrey's sheltering hypothesis. This observation agrees with our results. However, the phase shift reported in different experiments are usually in the $(\pi/2, \pi)$ range \citep[][]{DONELAN2006,GRARE2009}. 


To summarise, the $90^{\circ}$ phase shift, together with the pressure distribution strongly supports Jeffrey's sheltering hypothesis for the strongly forced waves ($c/u_*\leq 2$). This include both the non-separated cases for smaller $ak$ and intermittently separated cases for $ak$ above around 0.2. It is also where we see the smallest $S_p/S_{in}$ ratio, which indicates that the wave coherent viscous stress starts to play a role. The effect of viscous shear stress can be included in the sheltering parameter (\ref{eqn:Jeffrey_Sin}) as Jeffrey's original scaling analysis does not exclude the viscous shear stress. The transitional regime ($2\leq c/u_* \leq 4$) results in $\phi_{p1}\in (0, \pi/2)$. Only based on the phase shift, it does not seem to be explained by any existing theories, since both Miles's critical layer theory and Belcher's non-separated sheltering theory predict a negative $\alpha$. The other reason why Miles' critical layer theory does not apply to this regime is because the critical layer is very close to the water surface and affected by viscosity, therefore the inviscid assumption in Miles' theory does not hold. We note that the critical layer and the recirculating cells in the frame of reference of the wave still plays an important role in setting the pressure distribution, but does not necessarily follow Miles' calculation. Above the intermediate wave regime ($c/u_* \geq 8$), the phase shift $\phi_{p1}$ becomes slightly above $90^{\circ}$, which suggests that Miles’s critical layer theory and Belcher’s non-separated sheltering theory could potentially apply. 


\subsection{Notes on Reynolds number dependence}
A few processes discussed in the paper can be Reynolds number dependent (at least below some high asymptotic value). The transition to turbulence underwater is very likely sensitive to the Re number, together with the air side mean profile. The airflow separation is known to depend on the Re number, and consequently the phase shift of the principal mode of surface pressure, and the exact value of $F_p/\tau_0$  as well. Sensitivity to the Re number could contribute to the scatter observed in the wave form drag $F_p$ between numerical and experimental studies, although the transient nature of the wind wave growth problem and the effect of the highly nonlinear wave shape usually not characterized appear to already have a strong effect.

We argue that the most physically relevant Reynolds number we should use to cross-check different studies should be the one defined by the wavelength $Re_{\lambda}=u_*\lambda/\nu_a$ instead of $Re_*=u_*H/\nu$, since $Re_{\lambda}$ characterises the physically important ratio of length scales $k\delta_{\nu} = 2\pi/Re_{\lambda})$. The product of $k\delta_{\nu}$ and $c/u_*$ characterises the ratio of time scales $\omega t_{\nu}$ where $\omega=ck$ and the turbulence wall time scale $t_{\nu}=\delta_{\nu}/u_*$. For LES similarly, we should focus on the value of $k{z_0}$ where $z_0$ is the roughless length. In this way, the channel height is not relevant for the physics of the wind wave interaction (and ideally it should be at least a few wavelengths). The $Re_{\lambda}$ values we find in a few representative DNS works are: 130 in \citet{SULLIVAN2000}, 161 in \citet{KIHARA2007a}, and 283 in \citet{YANG2010}. For the present work it is 214. They are all on the same order of magnitude and therefore we cannot reach a definitive conclusion on whether the results are Reynolds number independent. The Reynolds number effects on the coupled wind-wave-current problem remain to be systematically investigated.

\section{Concluding remarks}
We have presented direct numerical simulations of wind waves forced by a turbulent boundary layer, by solving the two-phase Navier-Stokes equations. Leveraging these fully coupled and resolved two-phase DNS, we observe the complicated evolution of the fully coupled wind wave system, including the wave amplitude and shape change, the underwater drift current, and the feedback to the air side turbulent boundary layer.

Different from our previous study (2D laminar linear wind shear, small amplitude capillary gravity waves, and much lower $c/u_*$ ratio), the present work is centered around a different wind forcing mechanism more pertinent to the realistic finite amplitude gravity wave regime. We directly compare the wave energy growth against the pressure input and confirm pressure forcing as the major contribution to wave energy growth. We discuss the detailed pressure distribution (amplitude and phase) together with the integral quantities (drag force and energy input rate), for a wide range of wave steepness $ak$ and wave age $c/u_*$. The wave energy input rate is closely linked to the drag force and we discuss the scalings of the drag force and energy input rate with both $ak$ and $c/u_*$. Our results compare well to previous experimental and numerical works, while providing some possible explanations for discrepancies between different data sets.

The principal mode analysis on the surface pressure distribution feeds into the ongoing discussions on the exact mechanism responsible for wave growth under various wind forcing regimes. For the strongly forced case, the transient effect is important, and the pressure distribution agrees with the description of the sheltering effect proposed by Jeffrey, with airflow separation to some extent for the steeper cases. Miles' critical layer theory is not supported by the analysis on the pressure phase shift for $c/u_*<8$. We caution that some of the results might be Reynolds number dependent, which remains to be further studied.
 
We confirm that considering a prescribed wave shape and motion beneath a turbulent boundary layer is a reasonable approach for the weakly coupled cases (i.e. large wave age $c/u_*$ and very slow wave growth). We observe a good agreement between our results and previous numerical studies in this regime. However, in the strongly coupled cases (ie. small wave age and relatively fast wave growth), the transient nature of the problem leads to an evolution of the wave form drag, related to the evolving wave profile and short-crested wave shape, up to micro-breaking. This highlights the importance of a fully coupled approach for the strongly coupled cases. The current framework also opens great opportunities for studies of coupled air-water boundary layer, and breaking wind waves in the future.

\section*{Acknowledgements}
\label{sec:acknowledgements}
This work was supported by the National Science Foundation (Physical Oceanography) under Grant No. 1849762 to L.D., the High Meadows Environmental Institute Energy and Climate Grand Challenge and the Cooperative Institute for Earth System modeling between Princeton and the Geophysical Fluid Dynamics Laboratory (GFDL) NOAA. Computations were partially performed using the Extreme Science and Engineering Discovery Environment (XSEDE), which is supported by NSF Grant No. ACI-1053575; and on resources managed and supported by Princeton Research Computing, a consortium of groups led including the Princeton Institute for Computational Science and Engineering and the Office of Information Technology's High Performance Computing Center and Visualization Laboratory at Princeton University. J.W. would also like to thank the support of the Mary and Randall Hack ’69 Graduate Award received through the High Meadows Environmental Institute.

We declare no conflict of interest.

\appendix

\section{Mean profiles for different wave steepness and wave ages and the roughness length $z_0$}\label{sec:appendix0_2}
Here we present the mean wind velocity profile for cases of different initial $ak$ and $c/u_*$. A wave-fitted coordinate transform is defined when computing wave-averaged vertical profile (either the boundary layer underwater or the atmospheric boundary layer over waves) so that the region between the crest and the trough can be defined. The wave-following coordinate (denoted as $(\xi, \eta, \zeta)$) is obtained through the following implicit mapping:
\begin{align}
\begin{bmatrix}
    x \\
    y \\
    z 
\end{bmatrix} = 
\begin{bmatrix}
    x (\xi, \zeta) \\
    y(\eta) \\
    z (\xi, \zeta)
\end{bmatrix} = 
\begin{bmatrix}
    \xi \\
    \eta \\
    \zeta + h_w\cos(k\xi)\exp(-k|\zeta|)
\end{bmatrix} 
\end{align}
In the transformed coordinate, $\zeta=0$ corresponds to $z=h_w$.
Notice that this transformation only affects the area very close to the wave surface (say below $ k\zeta = 0.5$).

\begin{figure}
\vspace{0.5cm}
    \centering
    \includegraphics{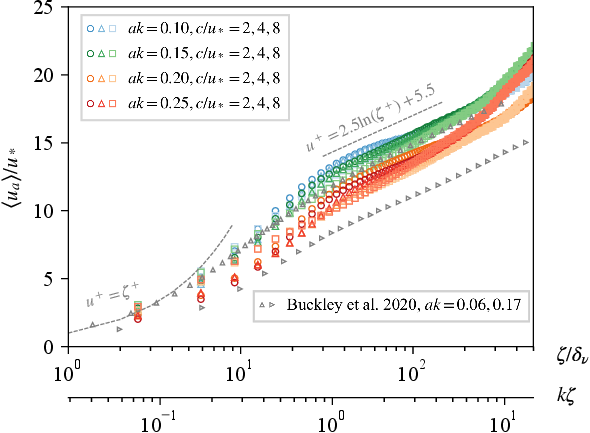}
    \caption{Mean wind velocity profiles for different wave steepness values and different wave ages. Generally there is a downshift of the profile at higher initial $ak$. Different shades of the same colour (and different symbols) represent different $c/u_*$, from dark to medium to light being $c/u_*=$ 2, 4, and 8. Plotted in triangles are the experimental results from \citet{BUCKLEY2020a}.}
    \label{fig:mean_allcases}
\end{figure}

Figure \ref{fig:mean_allcases} shows that the mean profiles resemble a typical linear-log profile with some deviation. In the near wall region, the mean profiles fall below the linear $u_a^{+} = \zeta^{+}$ because of a fraction of the wall stress is sustained by the wave form drag, as opposed to only the viscous stress in the flat wall case. In the logarithmic region, there is a downshift of the logarithmic region from the typical flat wall case (denoted with dashed line) since the waves' effect is similar to the roughness elements. Conventionally, a roughness length is introduced to represent this downshift so that
\begin{equation}
	\overline{u}_a(z) = \frac{1}{\kappa}\ln(\zeta/z_0)
\end{equation}
In this case, $z_0$ is a fitted value to the log region of the mean profile.
In our simulation, $z_0$ is generally larger for larger initial $ak$, although it seems to saturate at $ak=0.25$. For a given initial $ak$, the downshift is higher for higher $c/u_*$, although the effect is typically confined below $k\zeta=\pi$.

The trend of increasing $z_0$ with increasing $ak$ is consistent with experimental results. However, we find that the $z_0$ value in our cases is typically smaller, and the mean profiles are less cleanly linear-log than in the experiments. It is hard to find experimental evidence that directly discuss the effect of wave age on the mean profile, since in most experimental works the wave age and the steepness are coupled with purely wind driven waves.

The discrepancies are most likely due to the Reynolds number difference between the DNS and the experiments.
There are potentially two major non-dimensional numbers (ratios of length scales) that matter for the scaling of $z_0$: one is the wave steepness $ak$ (or $a_{rms}k$), the other is $a/\delta_{\nu}$. A recent study \citep[]{GEVA2022} suggests that the latter is the determining factor in their set of experiments with young, rapidly growing waves. However, assuming both matter for the more general case $z_0 = f(ak, a/\delta_{\nu})$, and it is equivalent to $z_0 = f(ak, k\delta_{\nu})$, The second ratio, as we have discussed in the paper, is determined by the Reynolds number $Re_{\lambda}$, and limited in the current DNS. Future studies should focus on how $k\delta_{\nu}$ effects the downshift of the mean profile.

The wave form drag $F_p/\tau_0$ we discuss at length in the paper is correlated to but does not translate directly into the roughness length $z_0$. It is a measure of how much form drag the surface creates, and quantifies the partition of energy and momentum flux into the waves. The relationship of $F_p$ and the wind profile is still unclear and requires further study.

\section{An initially steeper breaking case with $ak=0.3$}  \label{sec:appendix0_1}
\begin{figure}
    \centering
    \vspace{0.5cm}
    \includegraphics{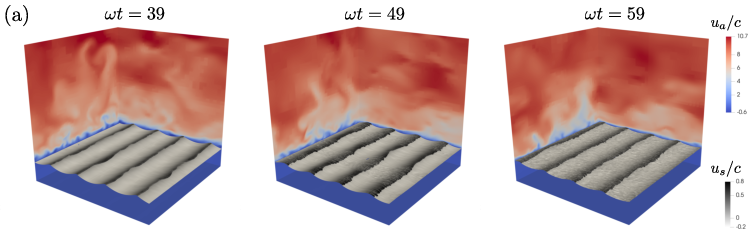}
    \includegraphics{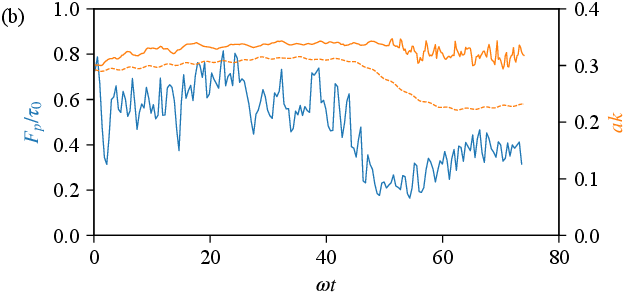}
    \caption{A breaking case with initial amplitude $ak=0.3$. The three frames show the waves and the wind before, during, and after breaking. The evolution of the $F_p$ as fraction of $\tau_0$, and wave steepness. There is a sharp drop of $F_p$ when the wave breaks around $\omega t  =0.4$. This again supports that the $F_p$ is mainly set by the wave steepness.}
    \label{fig:breaking_ak03}
\end{figure}
We have conducted a steep wave case (initial $ak=0.3$) which breaks within around 8 wave periods to demonstrate the solver's ability to simulate breaking waves with wind forcing. Figure \ref{fig:breaking_ak03} shows three frames around the breaking point. It resembles a typical spilling breaker with some droplets injection and rich 3D features. This breaking presents differences in terms of associated form drag compared to the micro-breaking described in \S\ref{sec:Fp_evolution} with initial $ak=0.2$ and long term wind forcing. The wave form drag decreases instead of increasing as in the micro-breaking case.

\section{Validation of the numerical method}  \label{sec:appendix1}

\subsection{Using adaptive mesh refinement in wall turbulence simulation}
In this study, we use Basilisk, a tree-based adaptive mesh refinement (AMR) solver to simulate a turbulent boundary layer flow. AMR exploits the fact that the dynamically active scales in the boundary layer is distributed inhomogeneously, and therefore the computation can be accelerated using a more refined grid near the wall and less refined grid away from the wall. Few works have applied AMR to the simulation of a turbulent boundary layer, as far as we know, except for \citet[][]{VANHOOFT2018} where AMR was used to perform large eddy simulation of the atmospheric boundary layer. We note that \citet{PERRARD2021,RIVIERE2021,FARSOIYA2021} have used AMR for an homogeneous and isotropic turbulence box and demonstrated the accuracy of the methods by considering the second order structure function scaling. 

\begin{figure}
\vspace{0.5cm}
    \centering
    \includegraphics{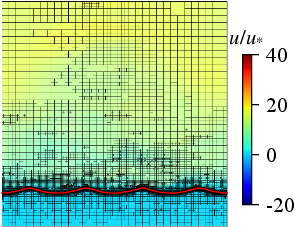}
    \caption{A slice of the field showing the adaptive mesh for the $ak=0.25$ case. The red curve is where the interface is. As we can see, the mesh is very refined around the interface.
    }
    \label{fig:mesh}
\end{figure}

Here, we directly solve the Navier-Stokes equation without any subgrid scale models, and we validate our approach against existing direct numerical simulation from \citet[][]{KIM1987} and verify that we reproduce the major features of the canonical turbulent wall-bounded flows. 

When simulating wall-bounded turbulent flows, the commonly adopted strategy to increase the near-wall resolution is to use (prescribed) non-uniformly spaced grid in the wall normal direction (e.g. Chebyshev grid in \citet[][]{KIM1987}), while keeping the spacing uniform in the streamwise and the spanwise directions. The adaptive mesh of Basilisk uses a different real-time adapting strategy based on the idea of wavelets. It was developed by \citet[][]{POPINET2003,POPINET2009}, with recent discussion in \citet[][]{POPINET2015} and \citet[][]{VANHOOFT2018}. Briefly speaking, once given the up-sampling ($U$) and down-sampling ($D$) operator (which are usually second-order) for computing a certain field ($f$) when the grid is refined and coarsened, the mesh is controlled by two parameters, the refinement criteria $\epsilon$ and the maximum level of refinement $N$. If the field is of size $L_0$, the smallest grid size is $\Delta = L_0/2^N$. For a given cell $i$ at level $n$, the discretization error is given by the absolute difference between the down-sampled and then up-sampled value and the original value \citep[][]{VANHOOFT2018},
\begin{equation}
    \chi_n^i = | U(D(f_n^i)) - f_n^i | 
\end{equation}
If $\chi_n^i$ is smaller that $2/3 \epsilon$, the $i$th grid is coarsened to level $n-1$; if $\chi_n^i$ is bigger that $\epsilon$, the $i$th grid is coarsened to level $n+1$ (only if $n+1 \ll N$); otherwise the $i$th grid is kept at level $n$.

In the simulation, we use an $\epsilon = 0.3u_*$ for the velocity field, and another $\epsilon_f = 10^{-4}$ for the volume fraction field $\mathcal{F}$. There can be fluctuations induced by AMR but the amplitude is directly controlled by the AMR refinement criteria. Since the AMR criteria are based on the velocity field rather than its spatial derivative (i.e. the deformation tensor used to compute the viscous stresses), the actual fluctuations on the stresses are not directly controlled by the AMR criteria. However, the numerical schemes (including the up-down sampling) are high-enough order (second order) that this should not affect the level of control on the stress fluctuations. The independence of the results on both spatial resolution and AMR thresholds has been checked, which includes the estimate of stresses.

\subsection{Comparison to canonical channel flow with $Re_*=180$}

To demonstrate that the turbulent boundary layer is resolved properly with the adaptive mesh, we perform a set of single phase channel flow simulations of $Re_{*} = 180$, and compare our results to the canonical DNS of a channel flow using a spectral method by \citet[][]{KIM1987}. In addition to validating our numerical method, the cases shown here also provide the benchmarks of how the controlling parameters of the adaptive mesh (i.e. refinement level $N$ and error tolerance $\epsilon$) affect the simulated flow.

\begin{table}
  \begin{center}
  \begin{tabular}{ccccccc}
  Case & $(L_x,L_y,L_z)/\delta$ &  \multicolumn{5}{c}{$\delta_{\nu}/\Delta$} \\
  \hline
  \citet[][]{KIM1987} ($Re_*=180$) & $(4\pi, 2\pi, 1)$ & \multicolumn{5}{c}{$z_1^+ = 20$*} \\
  \hline
  && N=7      & N=8     & N=9     & N=10    & N=11   \\\hline
  One-phase ($Re_*=180$) & $(2, 2, 1)$ & 0.36   & 0.71  & 1.42  &    &  \\ [3pt]
  One-phase ($Re_*=720$) & $(2, 2, 1)$ &    &   & 0.36  & 0.71   & 1.42 \\ [3pt]
  Two-phase ($Re_*=720$) & $(2\pi/(2\pi-1), 2\pi/(2\pi-1), 1)$ &   &  & 0.60   & 1.2   & 2.4  \\
  \hline
  \end{tabular}
  \caption{The number of grid points per viscous unit ($\delta_{\nu}/\Delta$) for different configurations and refinement levels. *The first grid spacing (often denoted as $z_1^+$) is not exactly comparable to the resolution in the AMR case: because stretched grid used in the spectral method, the grid size increases as it goes away from the wall.}
  \label{tab:N}
  \end{center}
\end{table}


\begin{figure}
    \centering
    \includegraphics{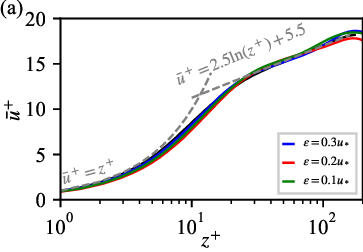}
    \includegraphics{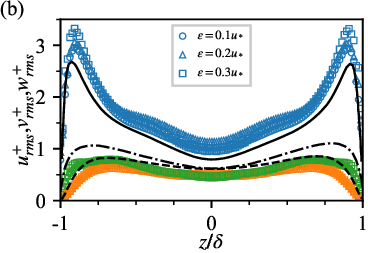}
    \caption{Turbulence statistics of one-phase channel flow with $N=9$. (a) Mean horizontal velocity in wall unit. $z^+ = z/\delta_{\nu}$; $\bar{u}^+ = \bar{u}/u_*$. Different colours represent cases of different error tolerances $\epsilon$. The black line is from \citet[][]{KIM1987}. (b) Velocity fluctuation. Blue: $u_{rms}^+ = u_{rms}/u_*$; green: $v_{rms}^+ = v_{rms}/u_*$; orange: $w_{rms}^+ = w_{rms}/u_*$ (wall normal velocity is $w$ in our coordinate system). Different marker shapes represent different error tolerances $\epsilon$. Black lines are from \citet[][]{KIM1987}. Solid line: $u_{rms}^+$; dash-dotted line: $v_{rms}^+$; dashed line: $w_{rms}^+$.
    }
    \label{fig:mean_rms}
\end{figure}

\begin{figure}
    \centering
    \includegraphics{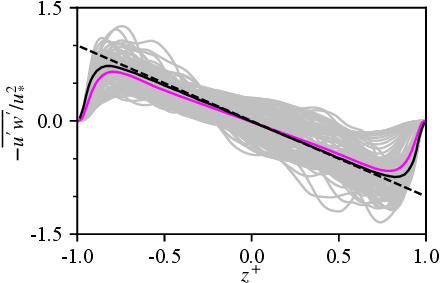}
    \caption{The Reynolds stress $-\overline{u^{\prime}w^{\prime}}$ normalised by total wall stress. The solid black line is from \citet[][]{KIM1987}. The computational domain in the AMR solver is by default cubed, and therefore limited in the streamwise and spanwise sizes. It causes the second order statistics to converge more slowly. Averaged over 10 eddy turnover time $T_e$, with $T_e$ defined as $T_e=\delta/u_*$.}
    \label{fig:Reynold_stress}
\end{figure}

The mean horizontal velocity $\bar{u}$ and the rms of velocity fluctuation $u_{rms}$, $v_{rms}$ and $w_{rms}$ are plotted in figure \ref{fig:mean_rms}(a) and (b) respectively. They both agree well with \citet[][]{KIM1987}, although there is a small difference in magnitude in the rms velocity. The mean profile converges at even very coarse grid spacing ($N=7$), which is an intriguing feature of AMR. The Reynolds stress shown by figure \ref{fig:Reynold_stress} also agrees with the reference case from \citet[][]{KIM1987}, despite taking longer to converge.

Notice that the refinement criteria $\epsilon$ has the same unit as the field $f$.
In the DNS of a turbulent channel flow case, we have found by trial and error that the $\epsilon$ value that works the best for the velocity field is around $0.3u_*$. It refines the near wall region without too much refinement in the centre of the channel. This is expected because the friction velocity $u_*$ is the characteristic velocity scale in the boundary, but we comment that the particular prefactor is likely to change for different configurations and Reynolds numbers. 

\subsection{Convergence between one-phase and two-phase cases at $Re_*=720$}

\begin{figure}
\vspace{0.5cm}
    \centering
    \includegraphics[scale=1]{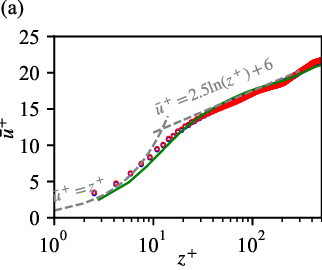}
    \includegraphics[scale=1]{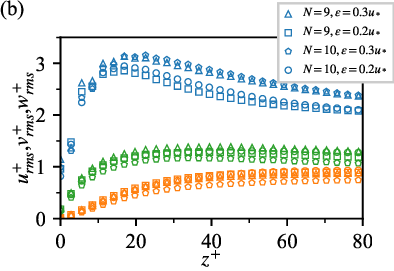}
    \caption{(a) Mean horizontal velocity for the $Re_{*}=720$ cases. Green curve: single phase with $N=9$, $\epsilon=0.3u_*$; red and blue dots: two phase cases with flat surface (the same configuration as all the moving wave cases), $\epsilon=0.1u_*$ at $N=10$ and 11 respectively. (b) The rms velocity for the single phase cases, under different maximum refinement levels $N$ and error tolerances $\epsilon$. Blue: $u_{rms}^+ = u_{rms}/u_*$; green: $v_{rms}^+ = v_{rms}/u_*$; orange: $w_{rms}^+ = w_{rms}/u_*$. }
    \label{fig:mean_720}
\end{figure}

The cases in the paper are run with the two-phase configuration at $N=10$ and $\epsilon=0.3u_*$ (see table \ref{tab:N}). We have also tested that the one-phase and two-phase flat wall cases agree with each other, and that the mean profile converges at $N=9,10,11$ (see figure \ref{fig:mean_720} (a)).
Figure \ref{fig:mean_720} (b) shows how the rms velocity is effected by the maximum refinement level $N$ and error tolerance $\epsilon$. A slightly larger $\epsilon$ results in  higher horizontal rms velocity in the outer region. Overall the difference is small and the rms velocity is well converged between different $N$ and $\epsilon$.

\subsection{Convergence verification for the moving wave cases}
\begin{figure}
\vspace{0.5cm}
    \centering
    \includegraphics{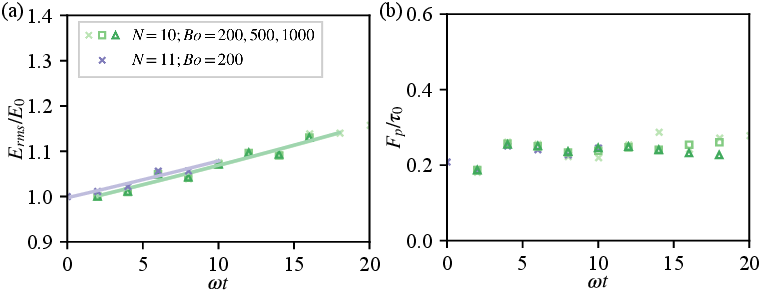}
    \caption{(a): convergence of the wave energy for different refinement levels $N$ and Bond numbers $Bo$. The energy evolution converges at higher Bond number. (b): convergence of the wave form drag. The symbols are the same with the left plot. $ak=0.15$, $c/u_*=2$.}
    \label{fig:convergence}
\end{figure}

We verify that the wave averaged quantities (energy and wave form drag) exhibit good convergence between the $N=10$ and $11$ cases, as we show in figure \ref{fig:convergence}. The results are also not sensitive when the Bond number is increased, as shown with different shades of green, confirming that the results in the paper apply in the gravity-capillary to gravity wave regime. Some variations in the wave form drag are seen, related to the chaotic variations of the instantaneous flow. 

\bibliographystyle{jfm}
\bibliography{reference}

\end{document}